\documentstyle[12pt,epsf]{article}
\newcommand{\simge}{\hspace*{0.2em}\raisebox{0.5ex}{$>$}
     \hspace{-0.8em}\raisebox{-0.3em}{$\sim$}\hspace*{0.2em}}
\def\simle{\hspace*{0.2em}\raisebox{0.5ex}{$<$}
     \hspace{-0.8em}\raisebox{-0.3em}{$\sim$}\hspace*{0.2em}}
\newcommand{\beq}{\begin{equation}}
\newcommand{\eeq}{\end{equation}}
\newcommand{\beqa}{\begin{eqnarray}}
\newcommand{\eeqa}{\end{eqnarray}}

\newcommand{\ga}{g_A}
\newcommand{\gas}{g_A^2}
\newcommand{\fp}{F_\pi}
\newcommand{\fps}{F_\pi^2}
\newcommand{\mpi}{m_\pi}
\newcommand{\mpis}{m_\pi^2}
\def\pislash{ {\pi\hskip-0.6em /} }

\def\nopi{ {\rm EFT}(\pislash) }
\def\lamchi{ \Lambda_\chi }

\def\si{{}^1\kern-.14em S_0}
\def\siii{{}^3\kern-.14em S_1}
\def\piii{{}^3\kern-.14em P_1}
\def\diii{{}^3\kern-.14em D_1}

\begin{document}

\begin{titlepage}

\hfill{NT@UW-01-06}

\hfill{LBNL 47692}

\hfill{RBRC-186}

\vspace{0.5cm}

\begin{center}
{\Large\bf Towards a Perturbative Theory of Nuclear Forces}

\vspace{1.2cm}

{\bf S.R. Beane}$^{a}$,
{\bf P.F. Bedaque}$^{b}$,
{\bf M.J. Savage}$^{a,c}$,
and 
{\bf U. van Kolck}$^{d,e}$

\vspace{1.2cm}
{\it
~$^a$ Department of Physics,
 University of Washington,\\
 Seattle, WA 98195\\
~\\$^b$ Nuclear Science Division,
Lawrence Berkeley National Laboratory,\\
Berkeley, CA 94720\\
~\\$^c$ Jefferson Laboratory,
 12000 Jefferson Avenue,\\
Newport News, VA 23606\\
~\\$^d$ Department of Physics,
 University of Arizona,\\
 Tucson, AZ  85721\\
~\\$^e$ RIKEN-BNL Research Center,
 Brookhaven National Laboratory,\\
Upton, NY 11973}
\end{center}

\vspace{1cm}

\begin{abstract}
  
  We show that an expansion of nuclear forces about the chiral limit is
  formally consistent and is equivalent to KSW power counting in the $\si$
  channel and Weinberg power counting in the $\siii-\diii$ coupled channels.
  Numerical evidence suggests that this expansion converges. The feasibility of
  making contact between nuclear physics and lattice-QCD simulations is
  discussed.

\end{abstract}

\vspace{2cm}
\vfill
\end{titlepage}
\setcounter{page}{1}


\section{Introduction}

The last decade has seen significant progress toward the formulation of an
effective field theory (EFT) description of multi-nucleon systems (for recent
reviews see Ref.~\cite{Be00,birarev,PKMR,EKNWGM,wkshop}).  The importance of
uncovering such an EFT cannot be overstated as it is this theory alone that
will allow for rigorous calculations of both elastic and inelastic processes in
multi-nucleon systems in a framework consistent with the Standard Model of
strong and electroweak interactions. An inherent advantage of an EFT framework
is that the uncertainty associated with the computation of any given observable
can be estimated and controlled with the power-counting scheme that defines the
theory (along with the regulator).  The fundamental difficulty in formulating
power-counting schemes for multi-nucleon systems is that the interaction of two
or more nucleons near threshold is intrinsically nonperturbative. This
manifests itself through the presence of scattering lengths that are much
larger than one would naively expect from QCD, $\sim \Lambda_{\scriptstyle
  QCD}^{-1}$~\cite{We90}. For very low-momentum processes, such as
$np\rightarrow d\gamma$ in the nucleosynthesis region, a consistent and
converging EFT, $\nopi$, has been developed to describe strong and electroweak
processes in multi-nucleon systems~\cite{nopi}.  However, attempts to develop a
fully consistent theory to describe processes involving momenta larger than the
mass of the pion have so far been unsuccessful.

Weinberg's (W) original proposal~\cite{We90} for an EFT describing
multi-nucleon systems was to determine the nucleon-nucleon (NN) potentials
using the organizational principles of the well-established EFTs describing the
meson-sector and single-nucleon sector (chiral perturbation theory), and then
to insert these potentials into the Schr\"odinger equation to solve for NN
wavefunctions.  Observables are computed as matrix elements of operators (that
have their own chiral expansion in this EFT) between these wavefunctions.  W
power counting has been extensively and successfully developed during the past
decade to study processes in the few-nucleon systems (see, for example,
Ref.~\cite{birarev}).  This method requires numerical solution of the
Schr\"odinger equation and is similar in spirit to traditional nuclear physics
potential theory.  Unfortunately, there are formal inconsistencies in W power
counting~\cite{KSWa}, in particular, divergences that arise at leading order
(LO) in the chiral expansion cannot be absorbed by the LO operators.  Problems
persist at all orders in the chiral expansion, and the correspondence between
divergences and counterterms appears to be lost. This formal issue was
partially solved by Kaplan, Savage and Wise (KSW) who introduced a power
counting in which pions are treated perturbatively~\cite{KSWb}.  The NN
phase-shifts and mixing angle in the $\si$ and $\siii-\diii$ channels have been
computed at next-to-next-to-leading order (N$^2$LO) in the KSW expansion by
Fleming, Mehen and Stewart (FMS)~\cite{FMS} from which it can be concluded that
the KSW expansion converges slowly in the $\si$ channel and does not converge
at all in the $\siii-\diii$ coupled channels. FMS found that the part of the
scattering amplitude that survives in the chiral limit is large enough to
destroy convergence in the $\siii-\diii$ channels. This is in contrast with
converging numerical results found using W power counting in this
channel~\cite{ray,baum}.

Both W and KSW power countings have desirable features that one would hope
exist in the, yet to be constructed, EFT that describes multi-nucleon systems.
It is therefore important to consider two fundamental questions in this area:

\begin{enumerate}
\item {\it Where is W power counting consistent?}  It has been argued that
  there are fundamental problems with the chiral expansion {\it and} the
  momentum expansion resulting from W power counting~\cite{KSWa}.  In this
  paper we confirm that W power counting is formally inconsistent in the $\si$
  channel, but we do find that W power counting is consistent in the
  $\siii-\diii$ coupled channels. The arguments presented in Ref.~\cite{KSWa}
  were based on a perturbative analysis, whereas the correct renormalization of
  singular potentials is intrinsically nonperturbative~\cite{PP,FLS,kiddies}.
  As we will see the same is true in the $\siii-\diii$ coupled channels.
  
\item {\it Where is pion exchange perturbative?}  The analysis of FMS
  demonstrated that pions, or more explicitly, the Yukawa part of the NN
  potential and radiation pions, when treated perturbatively give rise to a
  converging expansion for the $\si$ scattering amplitude.  They also showed
  that one-pion-exchange (OPE) in the $\siii-\diii$ coupled channels is
  not perturbatively convergent. In this paper we confirm the result of
  Ref.~\cite{KSWa} that part of the pion contribution to scattering in the
  $\si$ channel requires perturbative treatment in order to obtain a
  consistent theory.  However, there is no such requirement in the
  $\siii-\diii$ coupled channels.
\end{enumerate}

In this paper we will argue, supported by numerical evidence, that an {\it
  expansion about the chiral limit} provides a systematic power counting for
multi-nucleon systems.  This expansion is found to be equivalent to KSW power
counting in the $\si$ channel and equivalent to W power counting in the
$\siii-\diii$ coupled channels, i.e. it selects only the desirable features of
both power countings.  To understand the physics behind this expansion, it is
sufficient to recall that in the $\siii-\diii$ coupled channels, OPE includes a
strong tensor component at short distances that persists in the chiral limit
and is large enough so that it must be summed to all orders, as required by W
power counting.  However, deviations from the chiral limit $V({r};
m_\pi)-V({r}; 0)$ can be treated perturbatively in all channels, and in fact,
such a perturbative expansion is required in the $\si$ channel but not in the
$\siii - \diii$ channel.

In section 2 we discuss the $\si$ channel.  In particular, we review the
problems with W power counting in this channel.  Using the {\it Standard Toy
  Model of Nuclear Physics} (the three-Yukawa model developed in
Ref.~\cite{KSa}), detailed in the Appendix, we show that the Yukawa part of OPE
can be treated perturbatively if the short-distance physics is properly
accounted for in the EFT.  We outline a method for improving the convergence of
the $\si$ phase-shift in the realistic case.  In section 3 we investigate the
$\siii-\diii$ coupled channels.  The LO amplitude in W power counting is
analytically renormalized and a comparison with numerical data is presented.
The NN phase-shifts are computed at high orders in perturbation theory to study
the convergence of the $V({r}; m_\pi)-V({r}; 0)$ expansion.  In section
4 we discuss possible scenarios for the light-quark mass dependence of the
deuteron binding energy and NN scattering amplitudes, and dimly illuminate the
long and winding road between nuclear physics and lattice QCD. We conclude
in section 5.

\section{Scattering in the $\si$-Channel}

NN scattering in the $\si$ channel has been studied extensively in EFT
investigations as it results from a relatively simple interaction between
nucleons.  OPE provides the long-distance Yukawa interaction and there is a
large short-distance interaction, affectionately known as the {\it hard-core}.
This structure makes the $\si$ channel a useful testing ground for new ideas.

The $\si$ channel has been investigated using both W and KSW power countings
with interesting results. First, calculations up to N$^2$LO in W power counting
have been performed~\cite{ray,baum} and reproduce the phase-shift data in
this channel and in many other channels very well for a range of ultra-violet
(UV) momentum cut offs.  However, there is a sensitivity to the momentum cut
off due to an inconsistency in W power counting~\cite{KSWa}.  Second,
calculations up to N$^2$LO in KSW power counting have been
performed~\cite{KSWb,FMS}, and are found to converge to the results of the
Nijmegen phase-shift analysis~\cite{Nijmegen}.  The convergence of the
expansion can be improved substantially by promoting the effective range from
NLO to LO in KSW power counting~\cite{KSa}, implemented via the dibaryon
field~\cite{KadiB,BScount} or the z-parameterization~\cite{zparam}.

\subsection{Problems with W Power Counting in the $\si$ Channel}

In W power counting the momentum-independent four-nucleon interaction and OPE
form the LO NN interaction.  The Schr\"odinger equation is used to determine
the LO phase-shifts for NN scattering from this LO interaction.  This procedure
corresponds to summing an infinite number of insertions of the LO interaction.
Naively, this would appear to be a sensible procedure, however, subtle problems
arise in solving the Schr\"odinger equation due to its bad UV-behavior and it
must be renormalized, as discussed in detail in Ref.~\cite{KSWa}.

\begin{figure}[!ht]
\centerline{{\epsfxsize=3.in \epsfbox{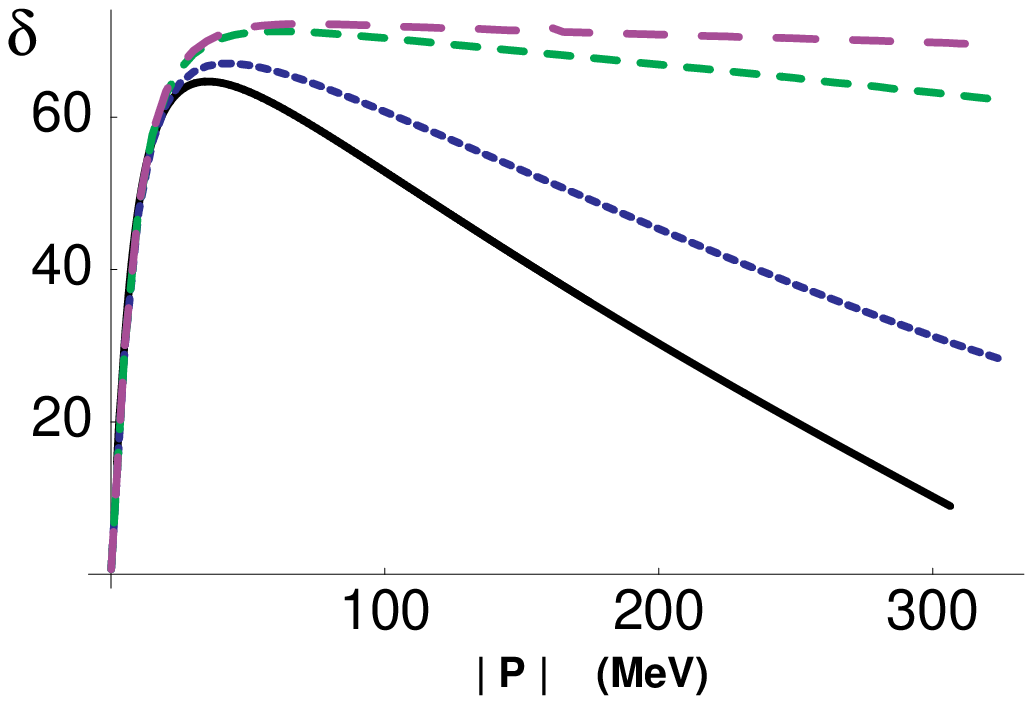}}
{\epsfxsize=3.in \epsfbox{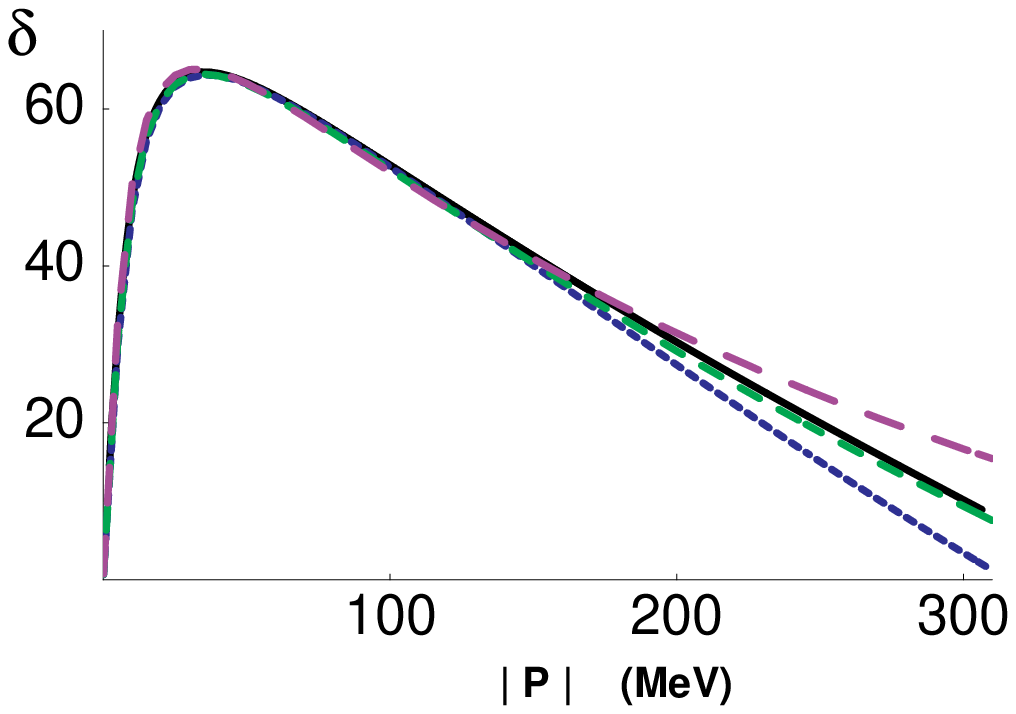}}} 
\vskip 0.15in
\noindent
\caption{\it 
The $\si$ phase-shift plotted versus CoM 
momentum. The solid lines are the Nijmegen phase-shift 
analysis~\protect\cite{Nijmegen}. 
The long-dash lines corresponds to $R=0.2~{\rm fm}$ 
($\Lambda=985~{\rm MeV}$), the medium-dash lines correspond to 
$R=0.4~{\rm fm}$ ($\Lambda=492~{\rm MeV}$), and the dotted lines 
correspond to $R=1.4~{\rm fm}$ ($\Lambda=140~{\rm MeV}$).
The left panel corresponds to setting $V_2=0$, while the right panel
includes $V_2$ in such a way to reproduce the measured effective range,
$r^{\scriptstyle\si}$.
}
\label{fig:singletW}
\vskip .2in
\end{figure}

The OPE central and tensor potentials are
\beq
V_C(r;\mpi) =-{\alpha_\pi}\,\mpis\;{e^{-\mpi r}\over r} 
\label{eq:central}
\eeq
and
\beq
V_T(r;\mpi) =-{\alpha_\pi}\,\mpis\;{e^{-\mpi r}\over r}
\left(1+{3\over{\mpi r}}
+{3\over{\mpis r^2}}\right)
\ \ ,
\label{eq:tensor}
\eeq 
respectively, where 
\beq {\alpha_\pi}={{\gas}\over{16\pi\fps}} 
\eeq 
and we use $\ga=1.26$, $\fp=93~{\rm MeV}$ and $M=940~{\rm MeV}$.  Scattering in
the $\si$ channel depends only on the central potential, $V_C$. Near the
origin, $V_C$ is Coulombic~\footnote{The $\delta^{(3)} (r)$ contribution to
  $V_C$ is not shown explicitly but is absorbed into the definition of the
  local four-nucleon operator defined in eq.~\ref{eq:ctovsinglet}.}  and gives
rise to NN wavefunctions that are a linear combination of regular and irregular
Bessel functions.  At LO in W power counting, OPE is renormalized by a single
delta-function counterterm, $C_0$.  For the purposes of this paper, we choose
to smear the delta-function counterterm(s) over a finite volume near the origin
and for computational ease choose a square well~\cite{scald} of radius $R$,

\beq {C_0}\;\delta^{(3)} (r)\rightarrow {{3{C_0}\; \theta (R-r)}\over{4\pi
    R^3}}\equiv {V_0}\;\theta (R-r).  \ \ \ 
\label{eq:ctovsinglet}
\eeq 
The OPE potential is similarly cutoff at the radius $R\equiv{1/\Lambda}$.  The advantage of
this simple regulator is that analytic matching can be performed~\footnote{ The
  physics we obtain will be regulator independent. Nevertheless, we have been
  unable to dimensionally regulate the $\siii-\diii$ coupled channels.}, and at
zero center-of-mass (CoM) energy one finds 

\beq {\sqrt{-M V_0}}\cot{\left({\sqrt{-M V_0}R}\right)}= 
-{m_\pi^2}M{\alpha_\pi}\log \left({R\over{R_*}}\right)
+{\cal O}(R)
\label{eq:logderivativeVsinglet}
\eeq
where $R_*$ is an intrinsic length scale to be determined numerically.
Treating ${\sqrt{-M V_0}R}$ as an expansion in $\mpi$, we find the depth of the
energy-independent square well as a function of $R$ to be

\beq
V_0(R;n)=-(2n+1)^2{{\pi^2}\over {4 M R^2}} -
{{2{m_\pi^2}{\alpha_\pi}}\over R}\log \left({R\over{R_*}}\right)
+{\cal O}(R^0).
\label{eq:logderivativeVsingletapp}
\eeq
\begin{figure}[!ht]
\vskip 0.15in
\centerline{{\epsfxsize=3.5in \epsfbox{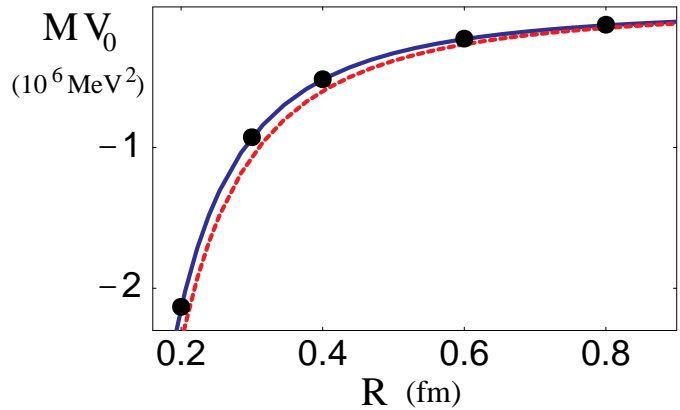}}} 
\vskip 0.05in
\noindent
\caption{\it 
The solid line represents the running of $V_0$ as a 
function of the cutoff $R$ (in fermis), taken from 
eq.~(\ref{eq:logderivativeVsingletapp}) with $n=1$, 
for the physical value of $m_\pi$, while
the dotted line neglects the $R^{-1}$ part of the running.
The dots are extracted directly from a 
numerical solution of the Schr\"odinger equation for the 
physical value of $m_\pi$.
}
\label{fig:vsinglerunning}
\vskip .2in
\end{figure}
where $n$ labels the branch of the cotangent. The presence of a
multi-branch structure is related to the accumulation of bound states
inside the square well. Of course the presence of unphysical bound
states is innocuous as long as the binding energies of such states 
are near the cutoff of the EFT.  
In  fig.~\ref{fig:vsinglerunning}
the $R$ dependence of $V_0$, as  given by
eq.~(\ref{eq:logderivativeVsingletapp}), 
is compared to points extracted from the numerical solution of 
the Schr\"odinger equation. 
For a given $R$, $V_0$ is tuned to give the scattering length,
$a^{\scriptstyle\si}=-23.714~{\rm fm}$, and in fig.~\ref{fig:singletW} we show
the $\si$ phase-shifts for different values of $R$. 
In momentum-space notation the matching equation takes the
form:
\beq
C_0(\Lambda;n)+\mpis D_2(\Lambda )=(2n+1)^2{{\pi^3}\over {3 M \Lambda}} +
{{8\pi{m_\pi^2}{\alpha_\pi}}\over{3\Lambda^2}}\log\left({{\Lambda_*}\over\Lambda} \right)
\ \ \ ,
\label{eq:logderivativeVsingletappczero}
\eeq 
and therefore the $D_2$ operator, which is formally subleading in W power
counting, {\it must be promoted to LO}, in agreement with the perturbative
argument of Ref.~\cite{KSWa}. Although the logarithmic divergence is suppressed
by a power of $\Lambda$ compared to the leading term in
eq.~\ref{eq:logderivativeVsingletappczero}, it is a true divergence in physical
quantities which must be renormalized at leading order in W power counting (see
eq.~\ref{eq:badlog} below).  This offers a proof that in the $\si$ channel, an
expansion of the action to a given order in W power counting {\it does not}
consistently remove all cutoff dependence to that order~\cite{lepage}. Of
course the $D_2$ contribution is numerically small, as demonstrated by the
dotted curve in fig.~\ref{fig:vsinglerunning} which neglects the $\Lambda^{-2}$
corrections to the running.  This smallness explains why W power counting has
been found to work well in this channel over a moderate range of cut
offs~\cite{ray,lepage,baum,PKMR}.

\begin{figure}[!ht]
\centerline{\epsfxsize=3in \epsfbox{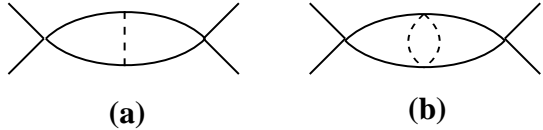}}
\vskip 0.15in
\noindent
\caption{\it 
Graphs with logarithmic divergences at (a) LO in W power counting 
and NLO in KSW counting and (b) NLO in W power counting
and N$^2$LO in KSW counting. 
The solid lines are nucleons and the dashed lines are pions.
}
\label{fig:wein_fig3}
\vskip .2in
\end{figure}

How does this problem manifest itself 
at the level of Feynman diagrams~\cite{KSWa}? 
In dimensional regularization the divergent part of diagram
(a) in fig.~\ref{fig:wein_fig3} is
\begin{eqnarray}
-{1\over \epsilon} {g_A^2 m_\pi^2 M^2\over 256\pi^2 F_\pi^2} C^2
\ \ \  ,
\label{eq:pole}
\end{eqnarray}
where $C$ is the coefficient of the four-nucleon operator, and $n=4-2\epsilon$
is the number of space-time dimensions.  This diagram requires a counterterm,
$D_2$, with a single insertion of the light quark mass matrix, $m_q$, to render
the amplitude finite, however, such counterterms occur only at higher orders in
the chiral expansion.  Thus, the inconsistency of W power counting is unmasked
diagrammatically.

%

It is instructive to understand how the divergence in eq.~(\ref{eq:pole})
arises in position space with a square-well regulator.  The radial
wavefunctions resulting from resumming the LO bubbles are phase-shifted
plane-waves of the form
\begin{eqnarray}
u(r) & \sim & \sin\left[ k r + \delta_{\rm LO}\right]
\ \ \ ,
\label{eq:LOpsi}
\end{eqnarray}
where $ \delta_{\rm LO}$ is the LO phase-shift.  Perturbatively
inserting the difference between the OPE potential
for $m_q\ne 0$ and its value in the chiral limit, and regulating the
UV divergence with a square well of a radius $R$ much smaller than the
scattering length $R\ll a$,  
the NLO contribution to the scattering amplitude is
\begin{eqnarray}
\langle \psi^\prime | V(m_\pi)-V(0) |\psi\rangle
& \sim & 
g_A^2 m_\pi^2
\int_R^\infty\ dr\ \sin^2\left[ k r + \delta_{\rm LO}\right]\ 
{\exp\left( - m_\pi r\right)\over r}
\nonumber\\
& \sim & 
-g_A^2 m_\pi^2  \sin^2\delta_{\rm LO}\  \log\left( m_\pi R\right)\ +\ ...
\ \ \ ,
\label{eq:badlog}
\end{eqnarray}
where the ellipses denote terms that are finite as $R\rightarrow 0$.  At the
order at which $ V(m_\pi)-V(0)$ enters, there {\it must} be a contribution from
a local operator of the form $m_\pi^2 \log\left(R/R_*\right)$, where $R_*$ is
some intrinsic length scale, in order to render the sum finite~\footnote{ If
  the LO radial wavefunctions behaved differently as $r\rightarrow 0$, the
  divergence structure would be modified. Such a modification can only arise
  from a re-ordering of the perturbative expansion, and consequently the
  hierarchy of higher-dimension operators as determined from naive arguments
  will also be modified.}  and independent of the renormalization scale $R$.

By the analysis above, one is naturally led to the conclusion that the
contribution from OPE and the ${\cal O}\left(m_q\right)$ counterterms must
contribute to any amplitude at the same order in the expansion, and this is
what leads to KSW power counting.  However, a more general conclusion to draw
is that the difference between the OPE contribution for $m_q\ne 0$ and the OPE
contribution in the chiral limit must occur at the same order as the ${\cal
  O}\left( m_q\right)$ counterterms.  In most cases, these two conclusions
yield identical amplitudes, however, in the $\siii-\diii$ channel they do not,
as we will see subsequently.

\subsection{Are Pions Perturbative?}

In KSW power counting the renormalization group (RG) flow of the LO
four-nucleon interaction promotes it to one lower order in the chiral
expansion, while the OPE interaction does not evolve.  The LO NN scattering
amplitude arises from resumming the bubbles with an infinite number of
insertions of the four-nucleon interaction.  At NLO, there are contributions
from OPE, insertions of ${\cal O}\left( m_q\right)$ four-nucleon operators and
from insertions of ${\cal O}\left(|{\bf p}|^2\right)$ operators.
Operationally, this hierarchy arises from the assignment that the scattering
length scales as an inverse power of the small expansion parameter, $a\sim
1/Q$, while all other parameters are natural size.  The scattering amplitude in
the $\si$ channel has been computed to N$^2$LO in KSW power counting in
Ref.~\cite{FMS}, where it was found to converge slowly. However, the cause of
the slow convergence with KSW in this channel stems from the treatment of the
short-distance contribution to the scattering amplitude and not from
perturbative OPE.  All momentum dependence arising from the hard-core of the NN
interaction is treated in perturbation theory and as these effects are large,
even over a relatively small momentum range, the convergence is slow.

\subsection{Lessons from a Model}

A toy model that has proved useful in understanding the structure of the NN
interaction is the three-Yukawa model~\footnote{We will refer to this model,
  which we detail in the Appendix, as the {\it Standard Toy Model of Nuclear
    Physics} (STMoNP).}.  One includes a Yukawa pion, rho and sigma meson, with
their physical masses (the sigma meson is given a mass of $550~{\rm MeV}$,
despite its non-existence in nature) and their couplings are adjusted to
reproduce the scattering length and effective range in the $\si$ channel.
Convergence problems with KSW power counting were highlighted with this model
by Steele and Kaplan~\cite{KSa}, and it is important to understand the cause of
these problems before proceeding further.

Several valuable lessons about both KSW and W power counting formulations can
be learned from the STMoNP.  First, pions are perturbative in the $\si$ channel.
The full theory defined via the Schr\"odinger equation can be systematically
perturbed in the number of insertions of the pion potential.  In addition, one
can recover these perturbative amplitudes with an EFT construction provided one
retains a sufficient number of counterterms.  Second, the fact that W power
counting~\footnote{In the STMoNP, W power counting does not lead to
  any ambiguities because the long-distance component of the nucleon
  interaction is sufficiently UV convergent.  } converges nicely, while KSW
converges quite slowly~\cite{KSa}, and appears to encounter problems at a scale
$\sim m_\pi$ is now understood.  The problem is not with the perturbative pions
component of KSW but rather with the treatment of the short-distance physics.
By resumming the ``hard-core'' into a dibaryon field, this modified-KSW (or
modified-W) power counting does provide a perturbatively-convergent amplitude.
The peculiar feature, however, is that by only including a small number of
local counterterms and by eyeballing the phase-shift plots, one would guess
that the breakdown scale for the theory is $\sim m_\pi$.  However, a more
systematic analysis, such as that provided by error plots or by going to
higher orders in the expansion, reveals a breakdown scale of $m_\sigma/2$, set
by the radius of convergence of the effective range expansion for the
short-distance physics.  Third, it is clear from this analysis that W power
counting is overkill.  One does not need to resum pions exchanges in order to
arrive at a meaningful phase-shift for this theory, and in fact, closed-form
analytic expressions exist and reproduce the numerical solutions of the
Schr\"odinger equation up to $m_\sigma/2$.

Note that the agreement with data (in the left panel of
fig.~\ref{fig:singletW}) breaks down at very-low momenta because the scattering
length is large and the effective range in this channel is large as well,
$r^{\scriptstyle\si}=2.73~{\rm fm}$, comparable to twice the pion Compton
wavelength.  This suggests that the two-derivative $C_2$ operator should be
promoted to LO, and is easily included in our analysis via an energy-dependent
contribution to the depth of the square well:

\beq
{C_2}\nabla^2\delta^{(3)} (r)\rightarrow 
{C_2}{M {\bar E}}\delta^{(3)} (r)\rightarrow 
{{3{C_2}k^2\;\theta (R-r)}\over{4\pi R^3}}\equiv k^2{V_2}\;\theta (R-r)
\eeq
where $\bar E=k^2/M$ is the CoM energy. The first step amounts to
a field redefinition which removes momentum dependence in favor of
energy dependence~\cite{BScount,harpaulo,RBM}. 
The value of $V_2$ as a function of $R$, is chosen to recover the experimental
value of $r^{\scriptstyle\si}$. In the right panel of fig.~\ref{fig:singletW}
we see the much-improved phase-shift with resummed range corrections, providing
an example of the manner in which the momentum expansion can be tailored to
experimental realities~\cite{BScount}.


\section{Scattering in the $\siii-\diii$ Coupled Channels}

In addition to the long-distance Yukawa OPE interaction and the 
hard-core interaction,
NN scattering in the $\siii-\diii$ coupled channels 
has a strong tensor component. 
At distances $r\ll\mpi$ the central potential is negligible
and the nucleons interact entirely through the singular tensor
interaction. 
The chiral expansion of the tensor force is equivalent to
the short-distance expansion and is given by
\beq
V_T(r;\mpi)\ =\ -{{3\alpha_\pi}\over{r^3}}\ +\ {{\mpis \alpha_\pi}\over{2r}}
\ +\ {\cal O}({m_\pi^4})
\ \ \ .
\label{eq:tensorchirallimit}
\eeq It is the leading term in this expansion which leads to the breakdown of
the KSW expansion in the $\siii-\diii$ coupled channels~\cite{FMS}.  Note that
a significant consequence of chiral symmetry is that there is no $1/{r^2}$ term
in the short-distance expansion of $V_T(r;\mpi)$.  One may hope that the
convergence problems in the $\siii-\diii$ coupled channels can be cured as in
the $\si$ channel by a re-ordering of local operators representing this short
distance physics.  Unfortunately, this is not the case.  For potentials of the
form $1/r^n$, with $n\leq 3$, $k\cot\delta$ is non-analytic in $k$ (for a
comprehensive review of singular potentials see Ref.~\cite{FLS}) and
consequently the chiral limit of $V_T(r;\mpi)$ cannot be represented by local
operators, and must be summed to all orders.  At first glance, this might seem
to pose an insurmountable problem for the EFT effort.  KSW have shown that if
one isolates terms in perturbation theory with insertions of the tensor force,
it appears that an infinite number of counterterms are necessary to properly
renormalize the highly-singular behavior. This was motivation for treating the
pion perturbatively.  Fortunately, this argument is not correct, essentially
because it relies on perturbation theory, and the correct renormalization is
invisible in perturbation theory. This is similar to what occurs in the
three-body system with short-ranged interactions~\cite{threebodies}.  In recent
work, some of us showed that a $1/{r^n}$ attractive singular potential can be
renormalized by a single energy-independent square well~\cite{kiddies}.  Here
we will extend this result to the realistic case, which amounts to showing that
the momentum expansion assumed by W power counting is consistent in the
$\siii-\diii$ coupled channels.  The investigation is in similar spirit to the
work in Ref.~\cite{FTT} who consider an OPE potential and work with a
subtracted T-matrix, and to the work in Ref.~\cite{dantom} which examines
scattering and electromagnetic observables in the $\siii-\diii$ coupled
channels.

\subsection{W Power Counting at LO}

At LO in W power counting, OPE in the $\siii-\diii$ coupled channels is
renormalized by a single delta-function counterterm, as it is in the
$\si$-channel.  For demonstrative purposes, we again choose to regulate the
delta-function with a square well, which has been implemented previously in
very forward-thinking work by Sprung {\it et al}~\cite{Sprung}. The $S$-wave
and $D$-wave components of the $J=1$ system, $u(r)$ and $w(r)$ respectively,
are coupled by the tensor potential, $V_T$, defined in eq.~(\ref{eq:tensor}).
The potential outside the square well is \beq {\cal V}_L(r;\mpi)= \left(
\begin{array}{cc}   
  -M V_C(r;\mpi)  & -2\sqrt{2}\; M V_T(r;\mpi)\\               
-2\sqrt{2}\; M V_T(r;\mpi) & -M\left( V_C(r;\mpi)-2 V_T(r;\mpi)\right)-{6/{r^2}}
          \end{array}
\right),
\eeq
while the potential inside the square well is
\beq
{\cal V}_S(r)=
\left(
\begin{array}{cc}   
-M (V_0 +k^2 V_2)  & 0\\               
0     & -M (V_0+k^2 V_2)-{6/{r^2}}
          \end{array}
        \right) \ \ .  
\eeq 
Here we have included $V_2$, but of course it is not necessary at LO in
W counting. We have also
left out $\siii-\diii$ superscripts which differentiate this square well from
that which appears in the $\si$ channel.  Defining $\Psi$ to be

\beq \Psi (r)= \left(\begin{array}{c}
            u(x)\\
            w(x)
            \end{array}\right)
          \ \ \ , 
\end{equation} 
the regulated Schr\"odinger equation is 
\beq
          {\Psi'' (r)}+ \left( k^2 + {\cal V}_L(r;\mpi)\theta (r-R) +{\cal
              V}_S(r)\theta (R-r)\right) \Psi=0.  
\end{equation} 
Renormalization concerns the behavior of physics at short-distances, or more
specifically, is due to our lack of understanding of physics at
short-distances.  If $R<r\ll M \alpha_\pi\equiv{1/\Lambda_{NN}}$, then we can
neglect the angular-momentum barrier and keep only the chiral limit of the
tensor force.  Moreover, for $|k|\ll\Lambda_{NN}$ the total energy can be
treated as a perturbation.  In this short-distance limit, ${\cal V}_L(r;\mpi)$
can be diagonalized and the Schr\"odinger equation can be solved exactly.  In
the diagonal basis the Schr\"odinger equation decouples into an attractive
singular potential and a repulsive potential. The solution for the attractive
singular potential, $\bar u$,  is a linear combination of Bessel
functions~\cite{Sprung,kiddies}, and it is useful to think of the
short-distance limit of the Bessel functions using the WKB
approximation~\cite{PP}.  The first two orders in the WKB approximation
reproduce the LO asymptotic expansion of the Bessel functions, while NLO in WKB
gives the leading energy corrections. The wavefunction at this order is

\beq
{\bar u}(r;k)= A\ r^{3/4}\cos\left( 2\sqrt{{{6 \alpha_\pi M}\over r}}
-{{r^2 k^2}\over 5}\sqrt{{r\over{6\alpha_\pi M}}}
 +\phi_0 + k^2 \phi_1\right),
\eeq
where $A$ is a dimensionful normalization constant
and $\phi_0$ and $\phi_1$ are the coefficients of a
Taylor-series expansion of the asymptotic phase $\phi (k^2)$.
Note that this wavefunction vanishes at the origin. 

\begin{figure}[!ht]
\centerline{{\epsfxsize=3.0in \epsfbox{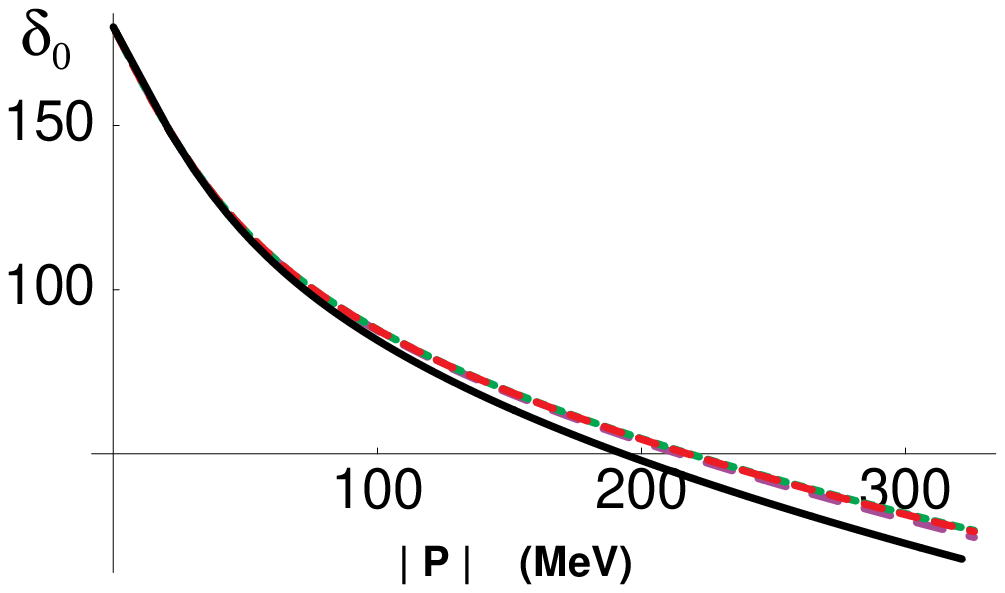}}
{\epsfxsize=3.0in\epsfbox{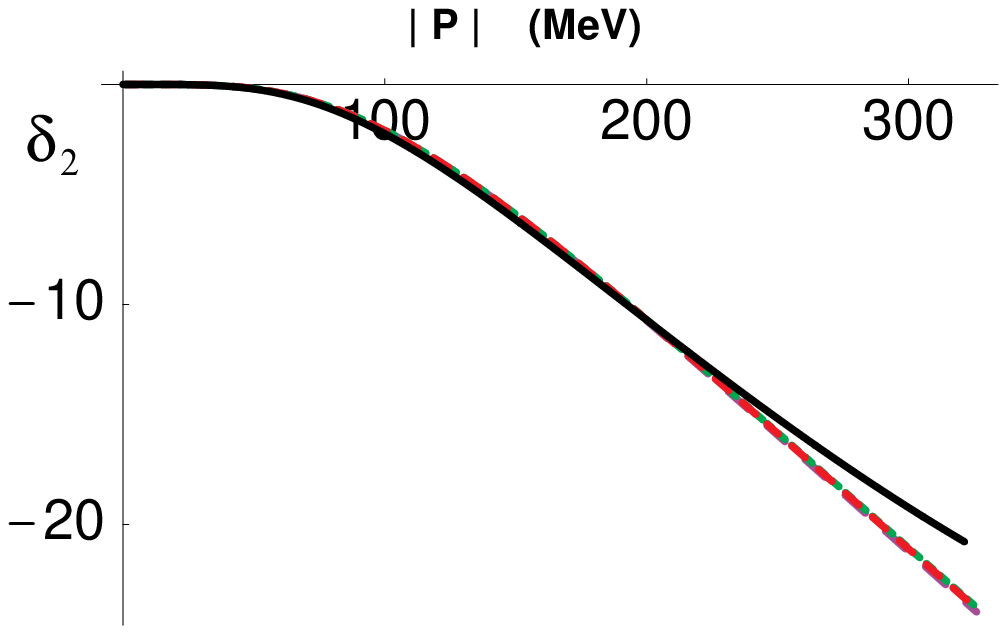}}} 
\vskip 0.15in
\noindent
\caption{\it 
The $\siii-\diii$ phase-shifts
as a function of the  CoM momentum. 
The solid line is the Nijmegen phase-shift analysis~\protect\cite{Nijmegen}. 
The long-dash line corresponds to $R=0.45~{\rm fm}$ ($\Lambda=438~{\rm MeV}$), 
the medium-dash line corresponds
to $R=0.21~{\rm fm}$ ($\Lambda=938~{\rm MeV}$),
and the dotted line corresponds to $R=0.10~{\rm fm}$ ($\Lambda=1970~{\rm MeV}$). 
The phase-shifts $\delta_0$ and $\delta_2$ are clearly cutoff-independent (see text).
}
\label{fig:delzeroW}
\vskip .2in
\end{figure}

\begin{figure}[!ht]
\centerline{{\epsfxsize=3.0in \epsfbox{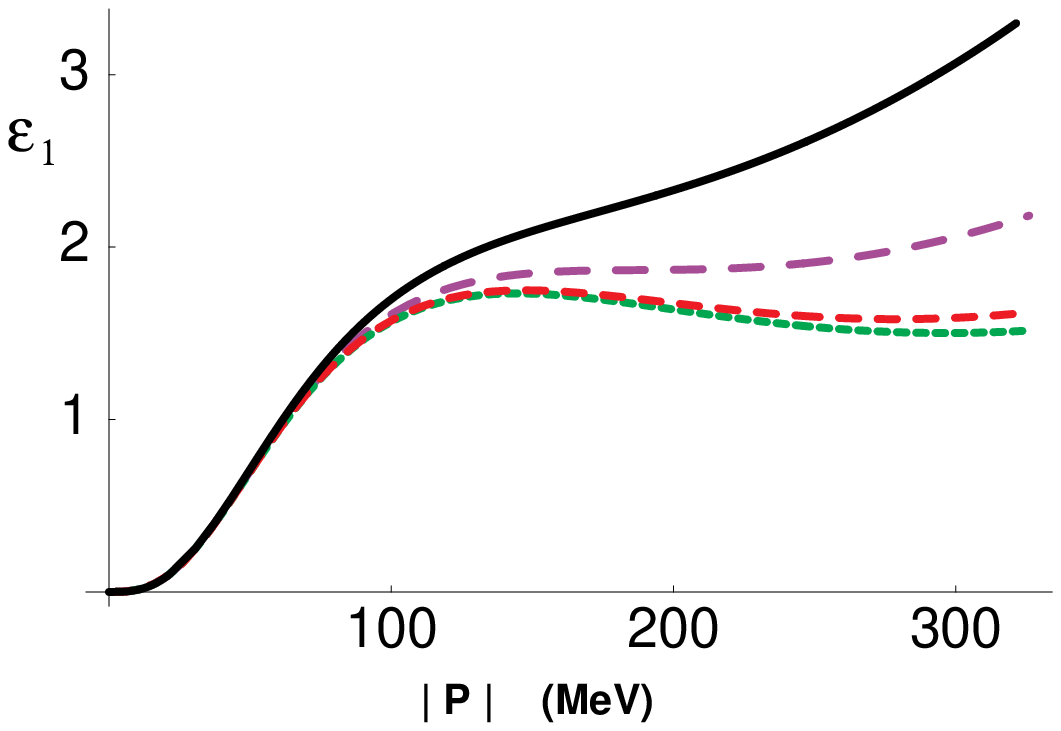}}
{\epsfxsize=3.0in \epsfbox{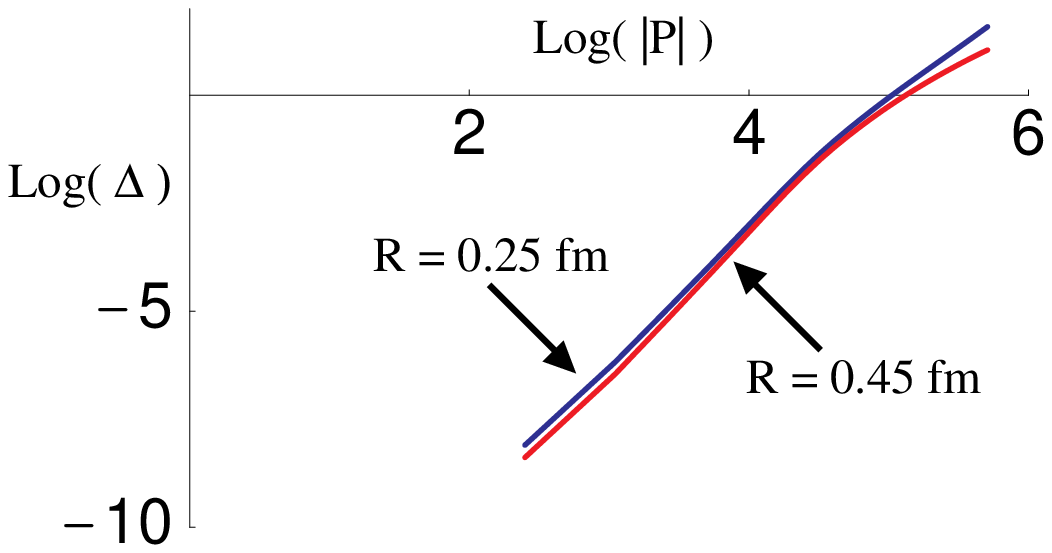}}} 
\vskip 0.15in
\noindent
\caption{\it 
The left panel shows $\varepsilon_1$ 
as a function of the  CoM momentum.. 
The solid line is the Nijmegen phase-shift analysis~\protect\cite{Nijmegen}. 
The long-dash line corresponds to $R=0.45~{\rm fm}$ ($\Lambda=438~{\rm MeV}$), 
the medium-dash line corresponds
to $R=0.21~{\rm fm}$ ($\Lambda=938~{\rm MeV}$), 
and the dotted line corresponds to $R=0.10~{\rm fm}$ ($\Lambda=1970~{\rm
  MeV}$). 
The right panel shows
an error plot for $\varepsilon_1$. Momentum $|p|$ is in units of ${\rm MeV}$. 
The quantity $\Delta$ is the difference
between the EFT calculation discussed in the text and the Nijmegen phase-shift
analysis. The slopes of the curves for $R=0.25~{\rm fm}$ and
$R=0.45~{\rm fm}$ are $\sim 2$.}
\label{fig:epW}
\vskip .2in
\end{figure}

The solution for $r<R$ is simply that of a square well of height
$V_0+k^2 V_2$.  Matching logarithmic derivatives of the interior
square well and exterior attractive solutions at
$r=R$ yields the matching equation
\beq
{\sqrt{-M V_0}R}\cot{\left({\sqrt{-M V_0}R}\right)}={3\over 4} + 
\sqrt{{6M{\alpha_\pi}}\over R}
\tan \left(2\sqrt{{6M{\alpha_\pi}}\over R}+\phi_0\right)
\label{eq:logderivativeV0}
\eeq
for $V_0$ and

\begin{eqnarray}
&&\qquad\quad{{(M V_2-1)}\over{M V_0}}\left[ \sqrt{-M V_0} 
\cot\left(\sqrt{-M V_0} R \right)
+ {M V_0}R \csc^2\left( \sqrt{-M V_0} R \right)\right]=\nonumber\\
&&{{R^{3/2}}\over\sqrt{{6 M \alpha_\pi}}}\tan \left(2\sqrt{{6M{\alpha_\pi}}\over R}+\phi_0\right)
-\left({2\over 5}R+{2\over R}\sqrt{{6M{\alpha_\pi}}\over
    R}\phi_1\right)\sec^2\left( 2\sqrt{{6 M \alpha_\pi}\over R}+ \phi_0\right)
\ \ \ \
\label{eq:logderivativeV1}
\end{eqnarray}
for $V_2$. These equations determine the renormalization-group flow of $V_0$
and $V_2$. Note that this flow is multibranched and non-analytic in the
coupling $\alpha_\pi$~\footnote{For sufficiently small coupling, analytic
  behavior in $\alpha_\pi$ is recovered, as detailed in Ref.~\cite{kiddies}.}.
The phases $\phi_0$ and $\phi_1$ are determined numerically.  For a given $R$,
$V_0$ is tuned to recover the physical value of the scattering length in the
$\siii$ channel, $a^{({\scriptstyle\siii})}=5.425~{\rm fm}$, from which
eq.~(\ref{eq:logderivativeV0}) determines the phase $\phi_0$.  $V_2$ is chosen
to recover the correct physical value of the effective range,
$r^{({\scriptstyle\siii})}=1.749~{\rm fm}$ from which
eq.~(\ref{eq:logderivativeV1}) determines $\phi_1$.
\begin{figure}[!ht]
\centerline{{\epsfxsize=3in \epsfbox{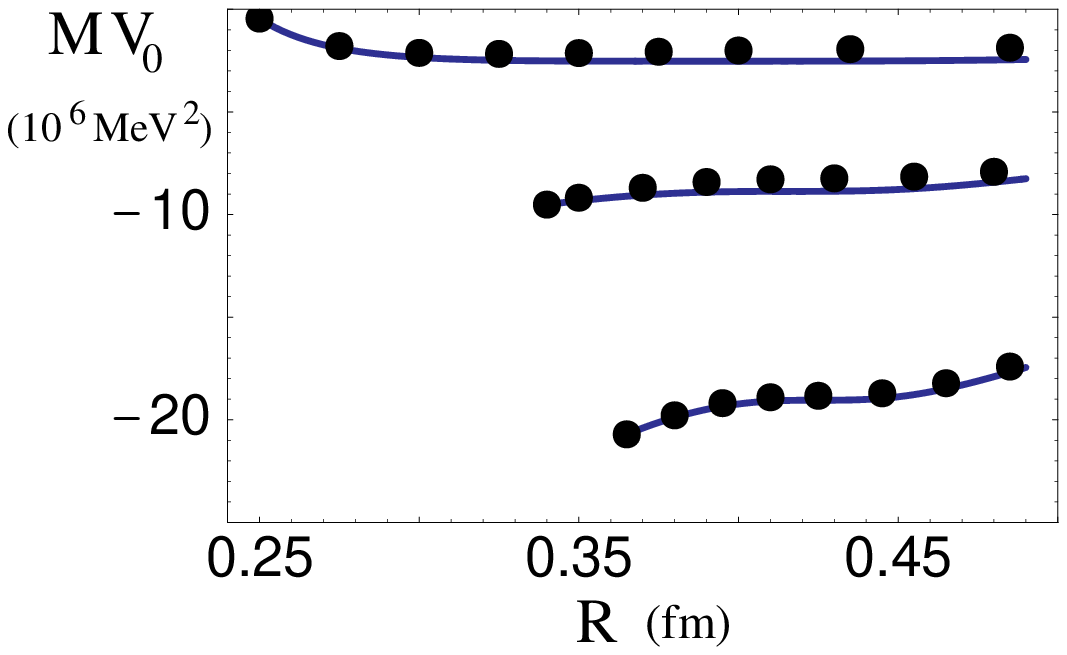}}\hskip0.1in{\epsfxsize=2.8in\epsfbox{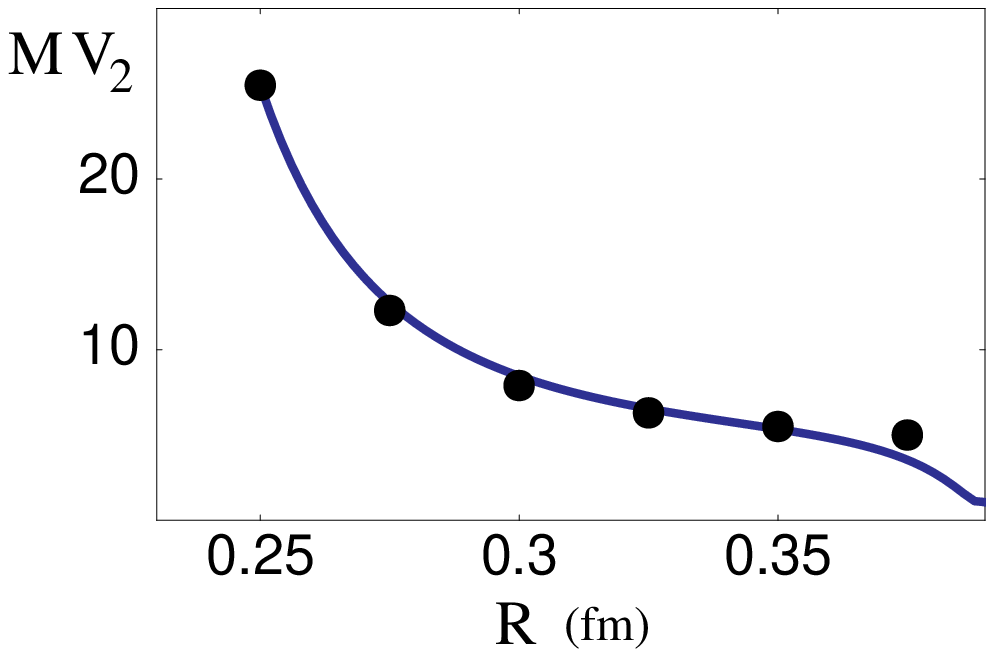}}} 
\vskip 0.15in
\noindent
\caption{\it The solid lines represent the running of $M V_0$ and 
$M V_2$ as a function of the 
cutoff $R$, taken from eq.~(\ref{eq:logderivativeV0}) 
and eq.~(\ref{eq:logderivativeV1}), respectively.
The dots are extracted directly from a numerical solution of the 
Schr\"odinger equation.
The different branches in the left panel correspond to a different number of nodes in the 
square well, i.e. an artifact of this particular regulator.
The curves continue to smaller values of $R$, but we have not shown them.
Further, analogous branches exist for $M V_2$, but we have not shown them.
For each branch, the ultraviolet phase is fit to the smallest $R$ data point to
produce the theoretical curve.
}
\label{fig:running}
\vskip .2in
\end{figure}
In fig.~\ref{fig:running} the square-well depths $V_0$ and $V_2$ are shown as
functions of the square-well width $R$ taken from
eq.~(\ref{eq:logderivativeV0}) and eq.~(\ref{eq:logderivativeV1}).  The LO
phase-shifts in the $\siii-\diii$ coupled channels are shown in
fig.~\ref{fig:delzeroW} and fig.~\ref{fig:epW} with $V_2=0$.  While the
phase-shifts $\delta_0$ and $\delta_2$ are clearly independent of the
square-well radius $R$, the mixing parameter $\varepsilon_1$ exhibits $R$
dependence.  However, an error plot of $\varepsilon_1$ indicates that the $R$
dependence and the deviations from the Nijmegen phase-shift analysis are higher
order in the momentum expansion, beginning at ${\cal O}\left( k^2 \right)$ (and
do not indicate an inconsistency in the power counting).

The scaling of the leading explicit chiral
symmetry breaking operators in the UV
can be found by perturbatively
inserting the difference between the OPE potential
for $m_q\ne 0$ and its value in the chiral limit,
\begin{eqnarray}
\langle \psi^\prime | V(m_\pi)-V(0) |\psi\rangle
 \sim  
g_A^2 {A^2}
\int_R^\infty\ dr\ r^{3\over 2}\cos^2
\left[2\sqrt{{{6 \alpha_\pi M}\over r}}\ \right]\ 
{{m_\pi^2}\over r}
 \sim  
-g_A^2 {A^2} m_\pi^2 R^{3\over 2} +\ ...
\ \ \ ,
\end{eqnarray}
where the ellipses denote terms that are finite as $R\rightarrow 0$.  From the
momentum-space perspective these leading explicit-breaking effects have a
strange $\lamchi^{-3/2}$ scaling where $\lamchi$ is the chiral symmetry
breaking scale.  In practice, $R$ is kept finite and while one might consider putting
in a $D_2$ operator in order to remove this $R$ dependence, it is formally not
necessary.  This is in contrast to the $\si$ channel, where logarithmic
dependence is found, eq.~(\ref{eq:badlog}), which does require the inclusion of
a $D_2$ operator.

\subsection{$\siii-\diii$ Phase-Shifts Perturbatively in 
$V({r}; m_\pi) - V({r};0)$}

\begin{figure}[!ht]
\vskip 0.15in
\centerline{{\epsfxsize=2.0in \epsfbox{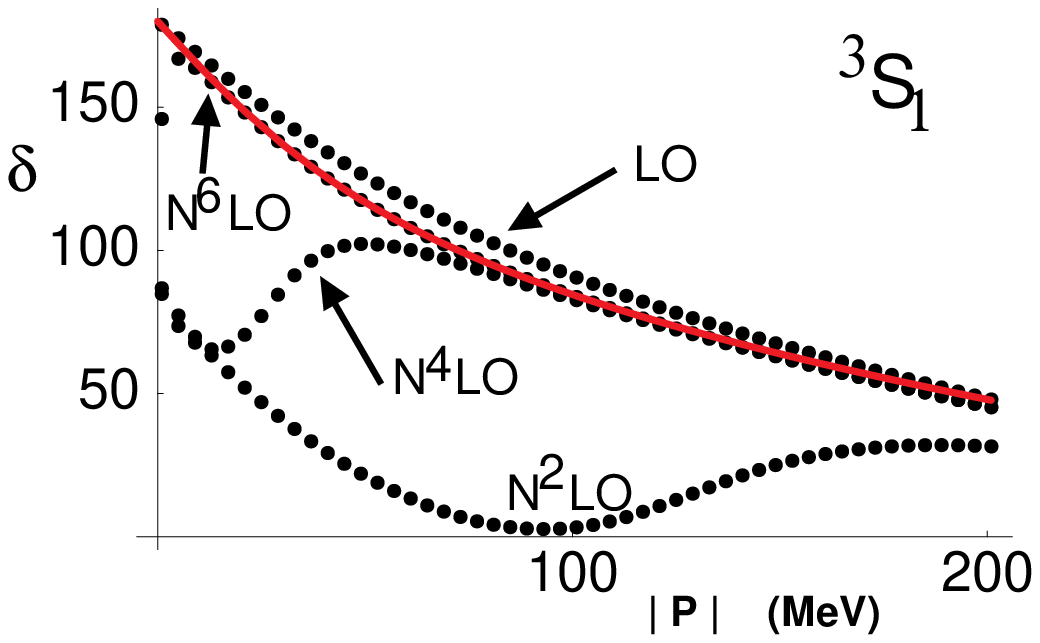}} 
{\epsfxsize=2.0in \epsfbox{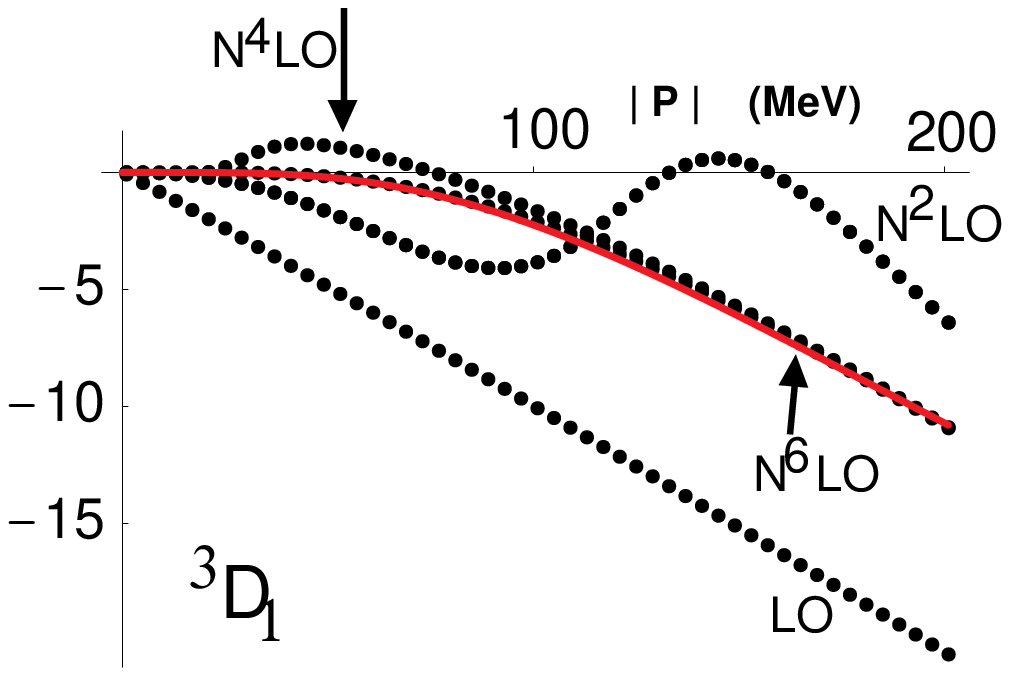}}
{\epsfxsize=2.0in \epsfbox{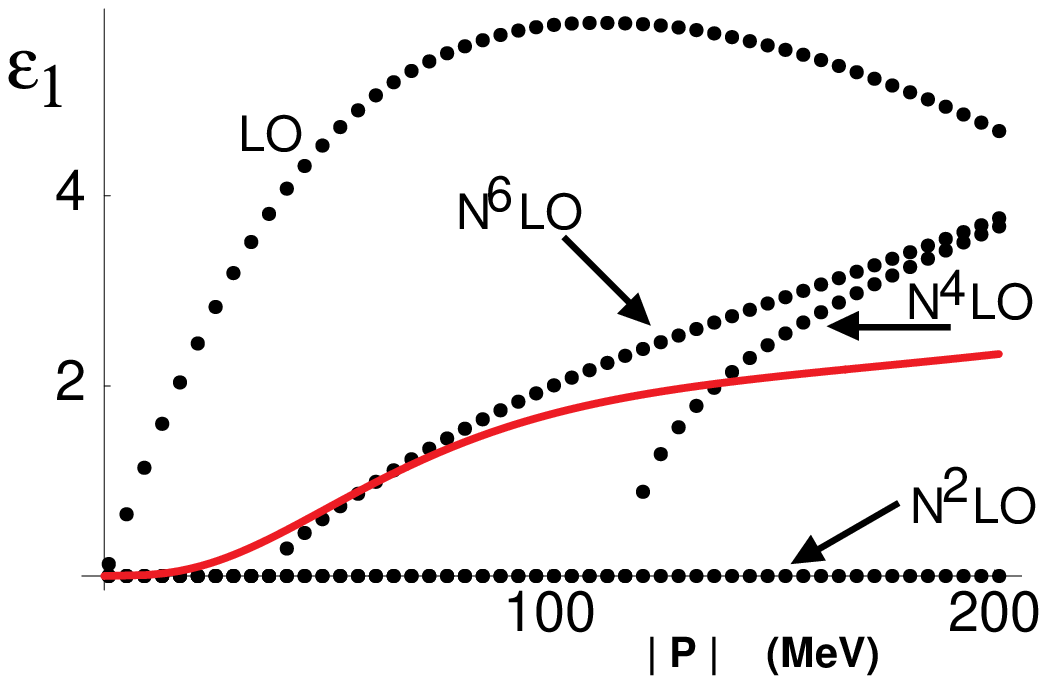}}}
\noindent
\caption{\it 
The $\delta_0$, $\delta_2$ phase-shifts and 
$\varepsilon_1$  mixing angle in perturbation theory out to N$^8$LO in
$V({r}; m_\pi) -V({r};0)$.
The solid line is the result of the Nijmegen phase-shift 
analysis~\protect\cite{Nijmegen}, 
while the dotted curves correspond to the LO, N$^2$LO, N$^4$LO and N$^6$LO
results, as indicated on each panel. We have not shown odd orders to
reduce clutter.
These results were obtained for a square well of
radius of $R=0.25~{\rm fm}$, 
with 
an energy-independent square-well depth of $M 
V_0 = 1.01\times 10^6~{\rm MeV}^2$ 
and 
$M V_2 k^2 = 23.54 k^2$. 
}
\label{fig:pertphases}
\vskip .2in
\end{figure}

There is now reason to believe that a perturbative expansion about the chiral
limit is formally consistent in both the $\si$ and $\siii-\diii$ channels.  In
order for this expansion to be useful it must also converge.  At NLO there will
be finite ${\cal O}(\mpis )$ corrections from the perturbative expansion in
$V({r}; m_\pi) - V({r};0)$.  However, since $\alpha_\pi$ is itself an expansion
about the chiral limit, the $D_2$ operator will contribute distinctly from the
$C_0$ operator. Hence, knowledge of $D_2$ requires a distinct measurement
involving multi-pion interactions with nucleons. In the absence of knowledge
about $D_2$, a well-defined, and significant computation that can be performed
is the perturbative expansion of the scattering amplitudes in $V({r}; m_\pi) -
V({r};0)$, the difference between the OPE potential evaluated at the physical
value of $m_\pi$ and the OPE potential for a massless pion.  Unfortunately,
this does not correspond to the chiral limit, as the couplings such as $g_A$
and $F_\pi$ are held fixed at their physical values~\footnote{ An expansion in
  $V({r};m_\pi)$ (not subtracting the $V({r}; 0)$) should recover the FMS
  analysis of KSW power counting at N$^2$LO~\cite{FMS} minus the contributions
  from radiation pions.}.  The STMoNP suggests that one should expect the
perturbative phase-shifts to be most deformed at lower-momenta, and approach
the observed value at higher momenta.  Fig.~\ref{fig:pertphases} shows the
phase-shifts and mixing angle of the $\siii-\diii$ coupled channels
order-by-order in perturbation theory out to N$^6$LO. The expansion is
convergent~\footnote{By convergent, we mean that the perturbative expansion
  converges to the solution to the Schr\"odinger equation for the full
  potential $V({r};m_\pi)$.}, but only slowly.
Despite the relatively poor convergence, this result is important as it
demonstrates that the expansion of the potential about the chiral limit is
convergent.  Clearly, one must be concerned about the slow convergence of this
expansion.  One hopes that this will not seriously impact the construction of a
systematic chiral expansion, but unfortunately it may.


\section{How Nuclear Forces Depend on $m_q$}

One of the important goals of the EFT investigations is to obtain the
dependence of nuclear observables upon the masses of the up and down quarks,
and eventually the strange quark.  This would allow one to determine the
deuteron binding energy in the chiral limit, and other interesting but perhaps
somewhat academic features of chiral-limit nuclear physics.  From a practical
point of view, knowledge of the quark-mass dependence of observables will allow
one to isolate terms in the chiral Lagrangian that contribute differently to
elastic and inelastic processes involving pions.  This is an important step
toward developing a unified theory of elastic and inelastic processes.
Further, knowledge of the quark-mass dependence of these observables will
provide the bridge between proposed lattice-gauge theory calculations,
performed at quark masses much larger than the physical values, and
nature~\footnote{ Recently, remarkable progress has been made in understanding
  how to compute higher-order coefficients of the chiral Lagrangian describing
  the pseudo-Goldstone bosons~\cite{Pqqcd}.  Partially-quenched lattice
  calculations have been matched onto a partially-quenched chiral Lagrangian
  and combinations of the Gasser-Leutwyler coefficients~\cite{GLthree}, the
  $L_i$'s, have been determined.  At this point in time, a partially-quenched
  EFT describing the single and multi-nucleon sectors does not exist.}.

In multi-nucleon sector calculations, quark-mass dependences are both
explicit and implicit.  For instance, in the NN-sector there is
explicit $m_q$-dependence in the mass of the pion in the nuclear
potential and also in four-nucleon operators.  However, there is
implicit $m_q$-dependence in the couplings in the meson sector
(e.g. $F_\pi$) and single-nucleon sector (e.g. $g_A$).  Without
measuring processes involving either initial or final-state pions, the
coefficient of each four-nucleon operator that has explicit $m_q$
dependence cannot be determined.  Therefore, what follows is for
illustration only, and we do not claim to presently know the deuteron
binding energy, or any other finely-tuned observable in the chiral
limit.

\subsection{Explicit $m_q$-Dependence only}

It is interesting to know the behavior of observables in nuclear
systems as a function of the explicit factors of $m_q$ that appear in
the nuclear potential.  That is to say, the value of the pion mass in
the OPE potential scales as $\sim m_q^{1/2}$ but other quantities, such
as $F_\pi$, $M$ and $g_A$ are fixed to their measured values.  It is
this scenario that can be directly compared with all potential-model
calculations, where the implicit $m_q$-dependence of the
short-distance part of the nuclear interaction is unknown.

\begin{figure}[!ht]
\vskip 0.15in
\centerline{{\epsfxsize=3in \epsfbox{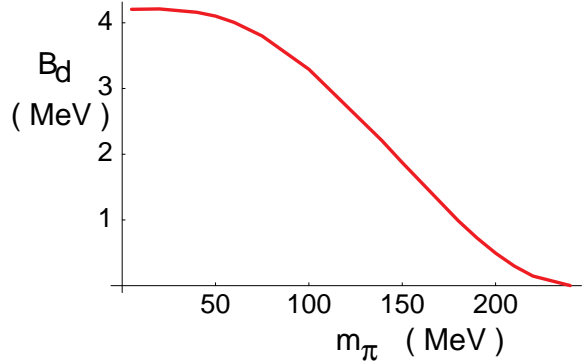}} }
\noindent
\caption{\it The deuteron binding energy as a function
of the $m_\pi$ that explicitly appears in the OPE potential.
For $R=0.25~{\rm fm}$, 
an energy-independent square well depth of $M 
V_0 = 1.01\times 10^6~{\rm MeV}^2$ 
and a square well that depends linearly on energy of
$M V_2 k^2 = 23.54 k^2$ reproduce the scattering length and
effective range in the $\siii$ channel for $m_\pi=139~{\rm MeV}$. 
}
\label{fig:Dbindpot}
\vskip .2in
\end{figure}

As we demonstrated earlier, with square-well regularization of the 
$\siii-\diii$ coupled channels, the $m_q$-dependent four-nucleon counterterm
required to yield a renormalization-scale ($R$) 
independent result scales like
$D_2\sim R^{3/2}$ as $R\rightarrow 0$, and therefore is naively higher order 
in the power counting.
Hence, we may vary the value of the explicit factors of 
$m_\pi$ in the nuclear potential without an accompanying modification to the
depth of the regulating square well.
A plot of the deuteron binding energy as a function of $m_\pi$ 
resulting from numerically solving the Schr\"odinger equation
is shown in fig.~\ref{fig:Dbindpot}.
For the physical value of the pion mass we recover the deuteron binding energy
to reasonable accuracy, $B_d=2.211~{\rm MeV}$ (essentially independent of $R$).
In the chiral limit one finds that the deuteron is bound by $B_d^0\sim 4.2~{\rm
  MeV}$.  This value is still somewhat small compared to $F_\pi^2/(2 M)\sim
10~{\rm MeV}$, that one might expect to arise in QCD, and therefore one would
conclude that the deuteron is still weakly bound in the chiral limit!  Our
calculation of the deuteron binding energy in the chiral limit agrees with that
obtained with the AV18 potential~\cite{BobW} of $B_d^{0,(AV18)}\sim 4.1~{\rm
  MeV}$. These very-similar results are in qualitative agreement (but of
opposite sign) with a previous estimate~\cite{BMS97} based on a first-order
perturbative calculation with wavefunctions from the Paris potential and Bonn
potential.

In addition to the deuteron binding energy, one gains useful insight
into the chiral limit of the NN-system by examining the phase-shifts and 
mixing parameters of the $\siii-\diii$ coupled channels.
We have considered the perturbative expansion of the phase-shifts and 
mixing angle previously, but it is worth restating in this discussion.
\begin{figure}[!ht]
\vskip 0.15in
\centerline{{\epsfxsize=2in \epsfbox{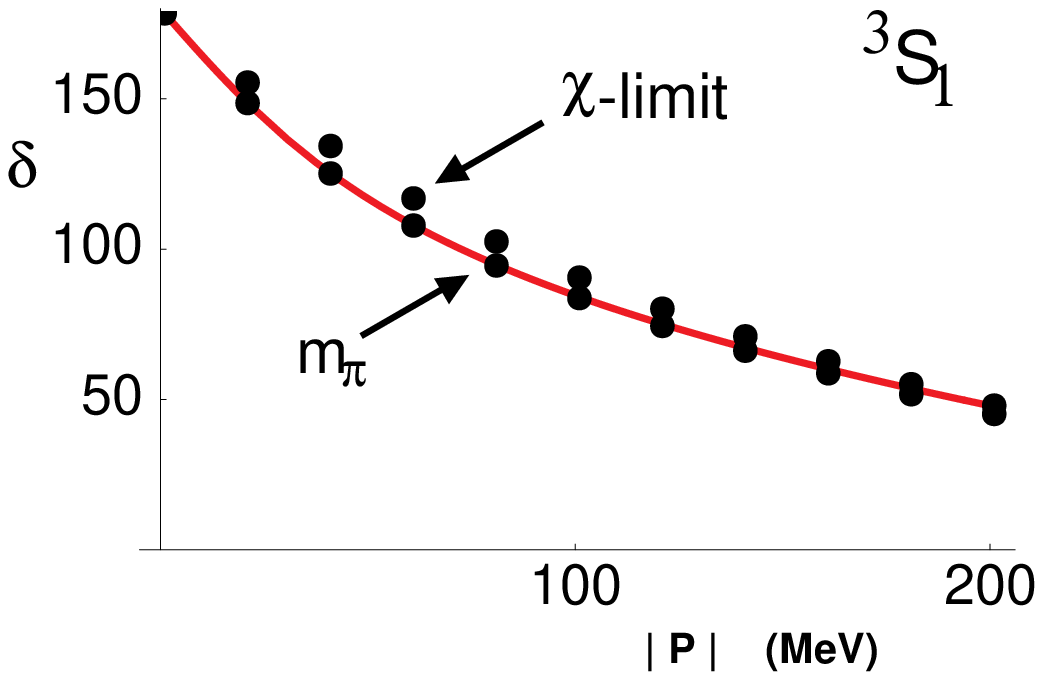}} 
{\epsfxsize=2in \epsfbox{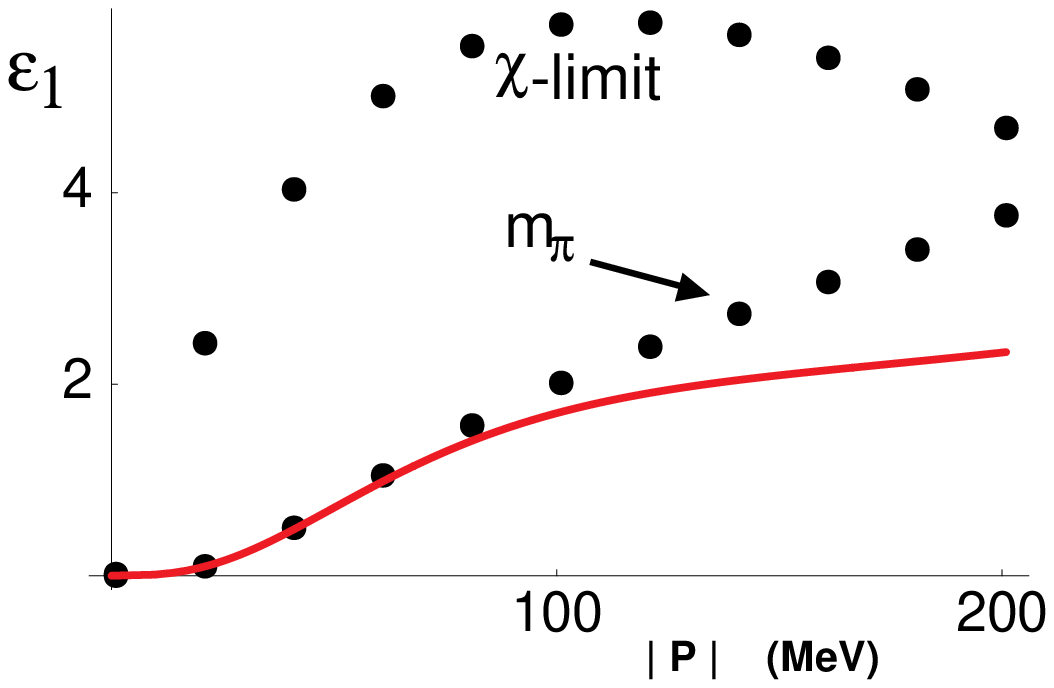}}
{\epsfxsize=2in \epsfbox{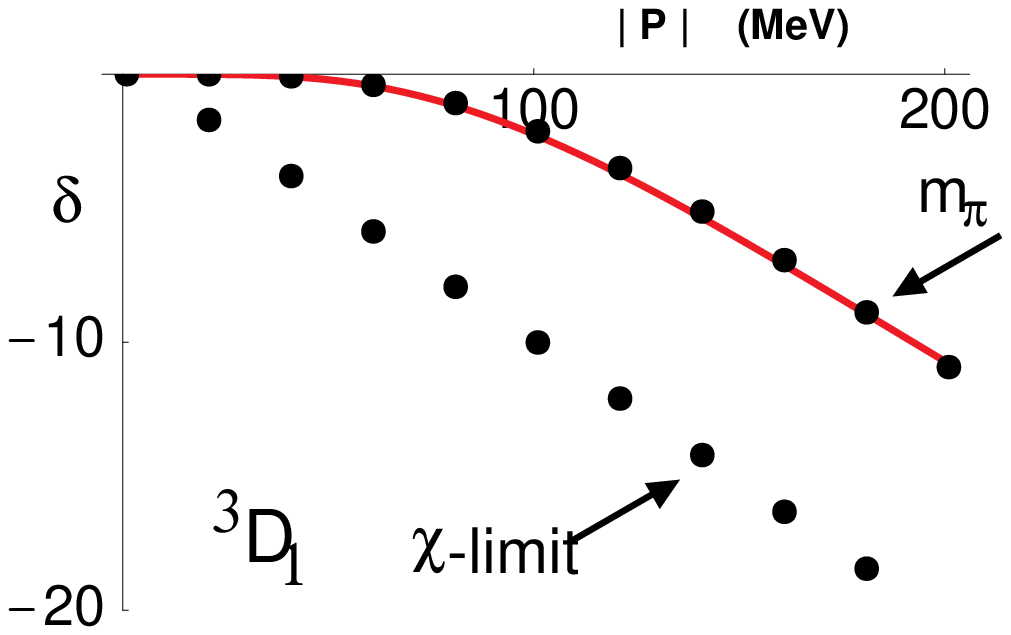}}
}
\noindent
\caption{\it 
The phase-shifts $\delta_0$, $\delta_2$ and the mixing parameter 
$\varepsilon_1$
of the $\siii-\diii$ coupled channels, generated by the parameter set 
as fig.~\protect\ref{fig:Dbindpot}.
The solid curves correspond to the Nijmegen phase-shift 
analysis~\protect\cite{Nijmegen},
while the two sets of dotted curves correspond to our calculated result
for the physical value of the pion mass, denoted by $m_\pi$ in the plots, and 
the chiral limit, as denoted in the plots.
}
\label{fig:SDphasepot}
\vskip .2in
\end{figure}
By numerically solving the Schr\"odinger equation we find the phase-shifts and 
mixing parameter shown in fig.~\ref{fig:SDphasepot}, see also
fig.~\ref{fig:pertphases}.
It is clear that the observed $\siii$ phase-shift does not differ 
significantly from that resulting from taking $m_\pi$ to zero in the potential.
However, as expected,
the $\diii$ phase-shift and mixing parameter $\varepsilon_1$ differ 
significantly from this ``chiral limit''.
As the deuteron is bound largely by the strong  attractive tensor force, 
the observed increase in $\varepsilon_1$ explains the deuteron binding energy 
computed above.

\subsection{Implicit and Explicit $m_q$-Dependence}

In addition to the explicit $m_q$ dependence of the OPE, there is implicit
$m_q$-dependence, as the couplings in the meson and single-nucleon sector also
depend upon $m_q$ at subleading orders.  Such dependences have been computed
previously~\cite{GaLe84} for QCD with $SU(2)_L\otimes SU(2)_R$ chiral symmetry.
The physical values of the pion decay constant $F_\pi$, the nucleon mass $M$
and the axial coupling constant $g_A$ are given in terms of the pion mass at
lowest order in the chiral expansion $m$, and $F$, $M_0$ and $g$, their
respective values in the chiral limit

\begin{eqnarray}
F_\pi & = & F \left[ 1 + { m^2\over 16\pi^2 F^2} \overline{l}_4\ +\ 
{\cal O}\left(m^4\right)\ \right]
\nonumber\\
M & = & M_0 - 4 m^2 c_1 + {\cal O}\left(m^3\right)
\nonumber\\
g_A & = & g \left[ 1 - 
{ 2 g^2+1\over 16\pi^2 F^2} m^2\log\left({m^2\over{\lambda^2}}\right)+{\cal O}(m^2)\right]
\ \ \ .
\label{eq:SNparams}
\end{eqnarray}
where $\overline{l}_4=4.4\pm 0.2$~\cite{GaLe84,CoGaLe01}, $c_1 \sim -1~{\rm
  GeV}^{-1}$~\cite{ulfioffe}, and we have use $m=140~{\rm MeV}$.  We have
retained only the leading chiral-logarithmic contribution to $g_A$, and have
chosen a renormalization scale of $\lambda=500~{\rm MeV}$.  In addition to
these relations, there is also $m_q$-dependence in the relation between the
$\pi NN$ coupling and the axial matrix element, as computed at LO in
Ref.~\cite{FeMe00},
\begin{eqnarray}
g_{\pi NN} & = & {g_A M\over F_\pi } 
\left( 1 - 2 {m_\pi^2\over g_A} \overline{d}_{18}\right)
\ \ \ ,
\end{eqnarray}
where we will use $\overline{d}_{18}\sim -1.0~{\rm GeV}^{-2}$~\cite{FeMe00}.
Even with this set of $m_q$-dependences included into the calculation, there
very well could
remain uncalculated $m_q$-dependences.
In particular, the implicit $m_q$-dependences modify the strength of the 
chiral-limit OPE potential.  This will give rise to a stronger R-dependence in
the local four-nucleon $m_q$-dependent operator with coefficient $D_2$.
Until processes involving external pions are computed, the contribution
of this operator cannot be isolated.
Therefore, for demonstrative purposes only, we do not modify the 
depth of the square wells evaluated for physical values of $m_\pi$ in our
chiral extrapolation.

\begin{figure}[!ht]
\vskip 0.15in
\centerline{{\epsfxsize=3in \epsfbox{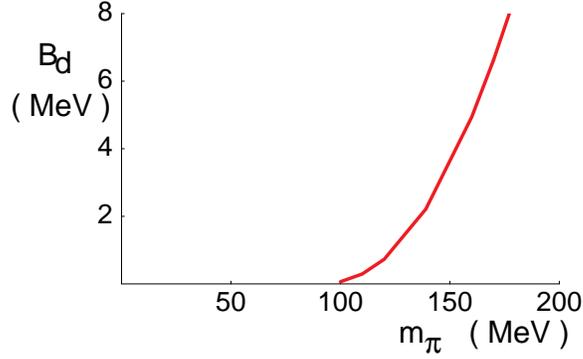}} }
\noindent
\caption{\it The deuteron binding energy as a function
of the pion mass that enters both explicitly and 
implicitly
in the nuclear potential.
For a radius of $R=0.25~{\rm fm}$, 
an energy-independent square-well depth of $M 
V_0 = 1.03\times 10^7~{\rm MeV}^2$ 
and a square well that depends linearly on energy of
$M V_2 k^2 = 15.25 k^2$ reproduce the scattering length and
effective range in the $\siii$ channel for the physical value 
of $m_\pi=139~{\rm MeV}$. 
}
\label{fig:DbindAll}
\vskip .2in
\end{figure}

The behavior of the deuteron binding energy as a function of
$m_\pi$ when both the implicit and explicit dependences are included
is shown in fig.~\ref{fig:DbindAll}.
In this scenario the deuteron binding energy decreases as the chiral limit is
approached, and the deuteron is unbound for $m_\pi\simle 100~{\rm MeV}$.  This
is precisely the opposite $m_\pi$-dependence to that found when only the
explicit dependence is retained.  Therefore, at this point in time a reliable
calculation of the $m_\pi$-dependence of the deuteron binding energy cannot be
performed.  It should come as no surprise that the fine-tuning in this channel
is significantly disturbed by what would appear to be higher-order
contributions.  Their impact is strongly enhanced by the large cancellation
between the kinetic and potential energy of the near-threshold bound state.

\begin{figure}[!ht]
\vskip 0.15in
\centerline{{\epsfxsize=2in \epsfbox{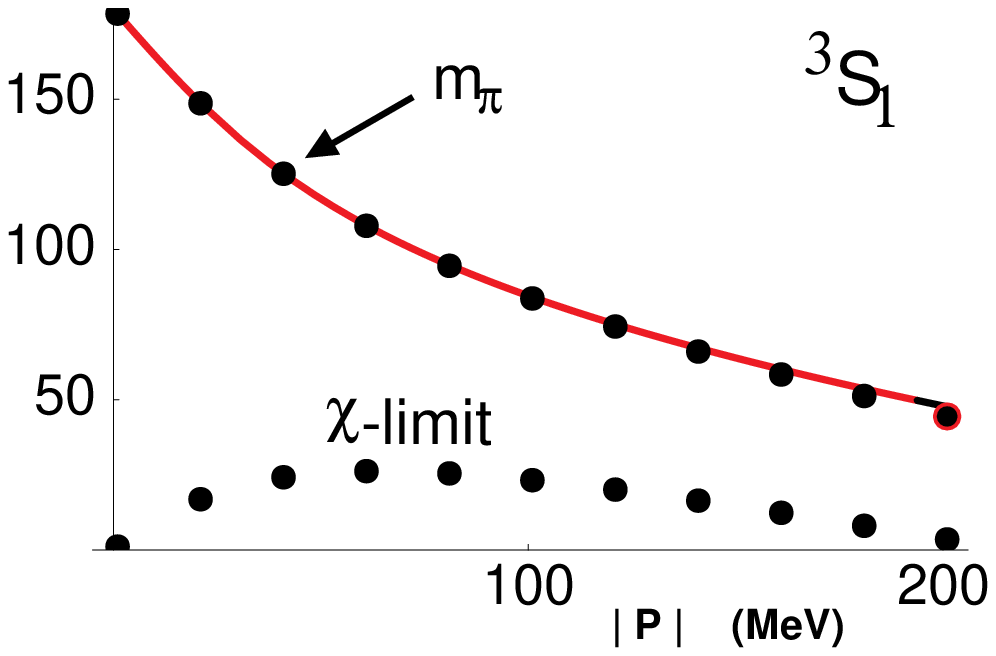}} 
{\epsfxsize=2in \epsfbox{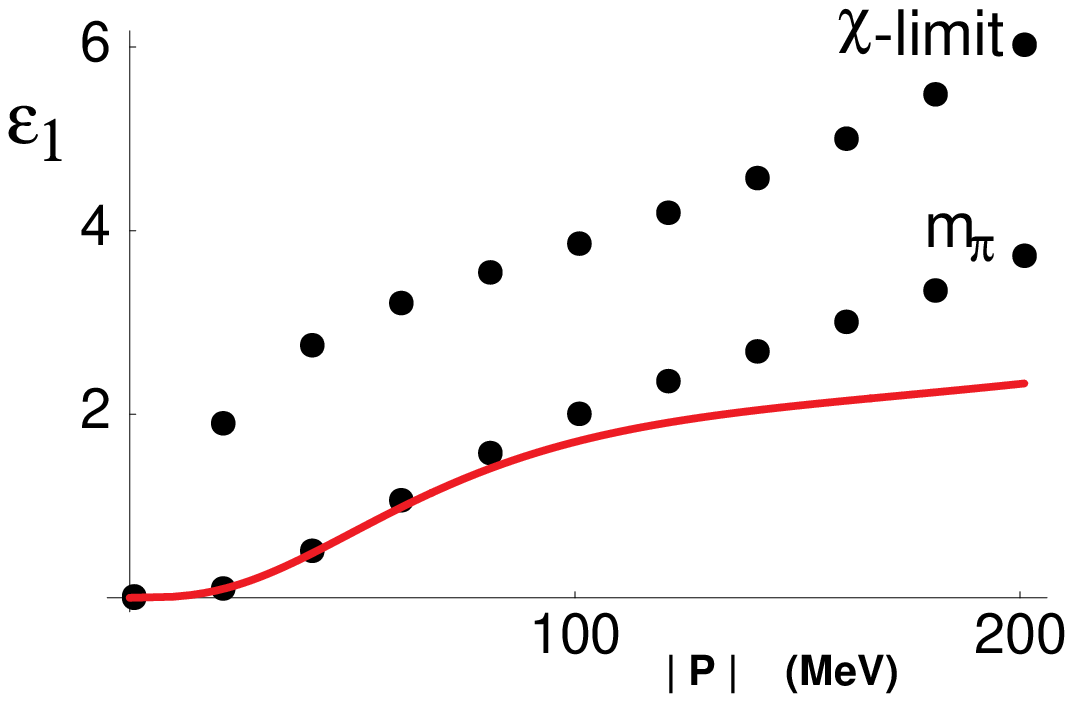}}
{\epsfxsize=2in \epsfbox{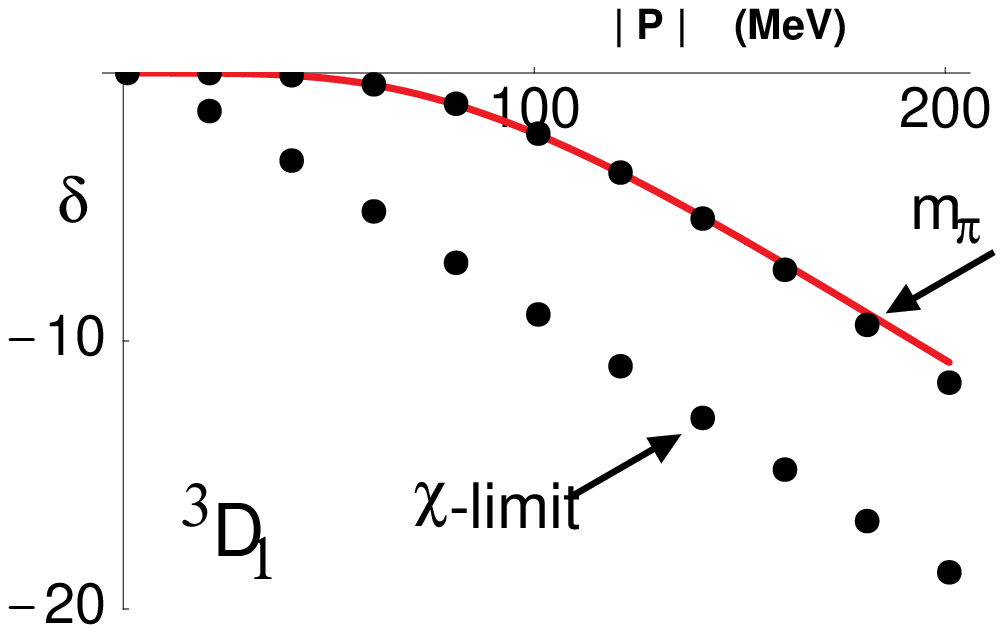}}
}
\noindent
\caption{\it 
The phase-shifts $\delta_0$, $\delta_2$ and the 
mixing parameter $\varepsilon_1$
of the $\siii-\diii$ coupled channels, 
generated by the parameter set 
as fig.~\protect\ref{fig:DbindAll}.
The solid curves correspond to the Nijmegen phase-shift 
analysis~\protect\cite{Nijmegen},
while the two sets of dotted curves correspond to our calculated result
for the physical value of the pion mass, denoted by $m_\pi$ in the plots, 
and the chiral limit, as denoted in the plots.
}
\label{fig:SDphaseAll}
\vskip .2in
\end{figure}

The $\siii-\diii$ phase-shifts and mixing parameter are 
shown in fig.~\ref{fig:SDphaseAll} for this scenario.
As required, the $\siii$ phase-shift is very different in this chiral
limit, due to the fact that there is no bound state.  On the other
hand, the $\diii$ phase-shift is very similar to that computed with
only explicit $m_\pi$-dependence retained.  Clearly, the
$m_q$-dependence of the OPE couplings makes a perturbatively small
contribution to this scattering amplitude, as one might expect (there
is no pole in this channel).

\subsection{Is an Expansion about the Chiral Limit Sensible?}

We have made the case that an EFT can be developed by performing an expansion
about the chiral limit.  Such an expansion overcomes the formal problem with W
power counting, that KSW power counting successfully resolved in the $\si$
channel.  However, given that the deuteron may be unbound in the chiral limit,
it is natural to ask whether such an expansion makes any sense.  The existence
or non-existence of the bound state depends upon subleading contributions to
the scattering amplitude.  That is to say, small changes in couplings have
large effects due to the inherent fine-tuning in the s-wave channels.

Therefore, in the $\siii-\diii$ coupled channels it may be consistent to
perform a chiral expansion of observables about a deuteron wavefunction defined
at high orders in the expansion, as has long been advocated by several groups
(see Ref.~\cite{We90,wavefunctionguys}) who have shown that many observables
are relatively insensitive to the choice of regularization.  For deuteron
observables, the EFT would then become an expansion of the interaction kernel,
up to the order at which the wavefunction is consistently defined.  Order by
order, observables would be independent of $R$ as $R\rightarrow 0$ if the
calculation is performed correctly.

\subsection{Matching to Numerical Results from Lattice-QCD}

In order to match to any foreseeable lattice-QCD results, an extrapolation in
$m_q$ will be required.  What we have presented in this work is the framework
required to match to a fully unquenched lattice computation, and we have shown
that there is currently no control of the extrapolation away from the physical
value of $m_\pi$, due to the fine-tunings that exist in nature.  However,
future lattice computations are likely to be partially-quenched~\cite{Pqqcd},
as opposed to unquenched.  Therefore, a partially-quenched EFT will need to be
constructed in order to correctly perform the $m_q$-extrapolation from the
lattice masses, to the physical masses.  Given our relatively primitive
understanding of the EFT for QCD, and also considering that the
partially-quenched EFT required to perform extrapolations in the single-nucleon
sector does not exist at this point in time, a significant theoretical effort
is required to put in place a framework that will enable future lattice-QCD
computations to make rigorous statements about multi-nucleon systems.

\begin{figure}[!ht]
\vskip 0.15in
\centerline{{\epsfxsize=3in \epsfbox{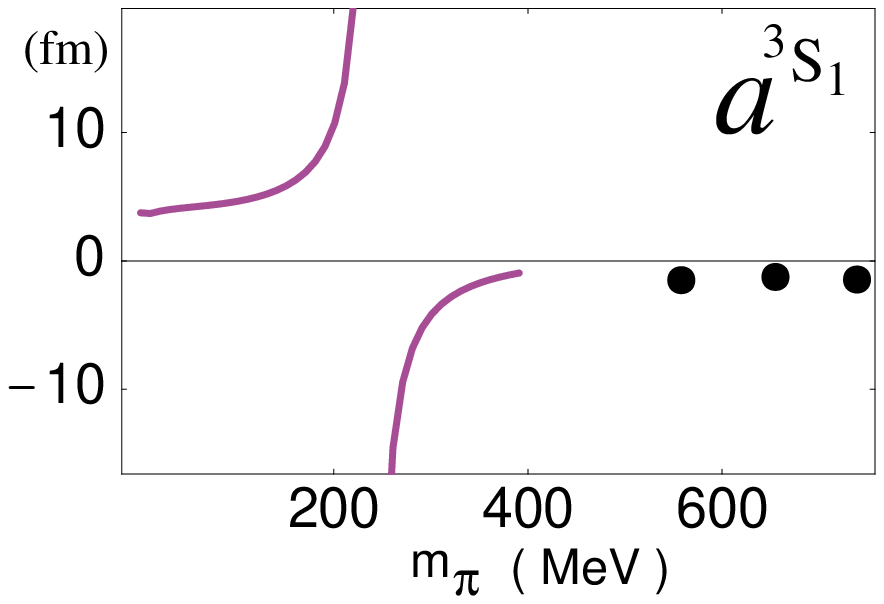}}
{\epsfxsize=3in \epsfbox{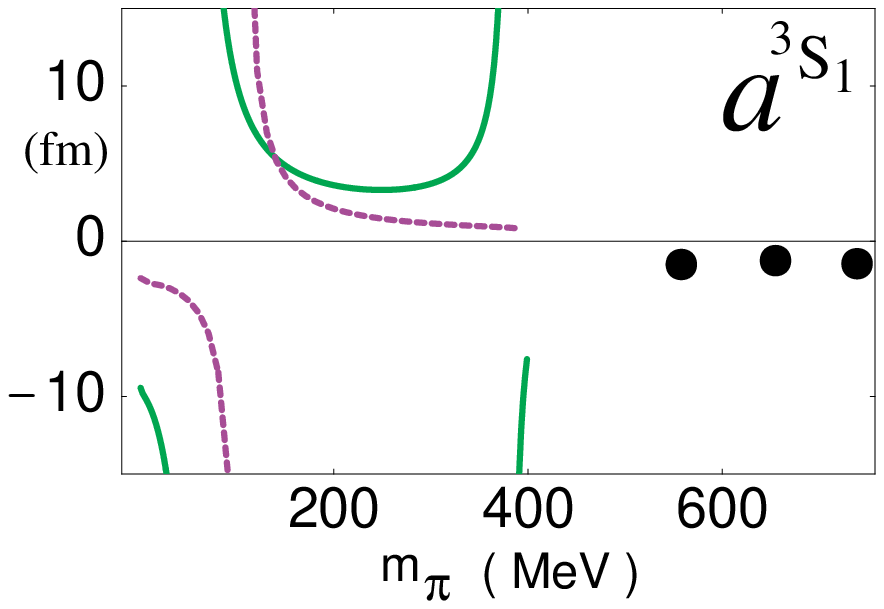}} }
\noindent
\caption{\it 
The $\siii$ scattering length as a function of $\mpi$. 
In the  left panel, the quantities $g_A$, $M$ and $F_\pi$ are
set equal to their physical values for all $m_\pi$.
In the right panel, the LO $m_\pi$ dependence of 
$g_A$, $M$ and $F_\pi$ is included.
The dashed (solid) line corresponds to a radius of
$R=0.25~{\rm fm}$, ($0.45{\rm fm}$) and an energy-independent
square-well depth of $M V_0 =-1.04\times 10^4 ~{\rm MeV}^2$
($-7.82\times 10^6{\rm MeV}^2$) which reproduce the scattering length
for the physical value of $m_\pi=139~{\rm MeV}$. 
The dots are lattice
QCD data, as explained in the text.}
\label{fig:scattpanel}
\vskip .2in
\end{figure}

It is exciting to note that attempts have been made to compute NN scattering
parameters with quenched lattice-QCD.  Ref.~\cite{fuku} computes the $\si$ and
$\siii$ scattering lengths in quenched QCD at $\beta=5.7$ with Wilson quarks
using L\"uscher's finite-volume algorithm, which expresses the energy of a
two-particle state as a perturbative expansion in the scattering length divided
by the size of the box~\cite{Luscher}. Practical computations require a
sufficiently small box which in turn should be larger than the scattering
length. Hence L\"uscher's method is ideal for systems with ``natural''
scattering lengths, that is with size of order a characteristic physical length
scale. Of course in the NN system, the scattering lengths are much larger than
the characteristic physical length scale given by $\Lambda_{\scriptstyle
  QCD}^{-1}$. Nevertheless one might suppose that away from the physical value
of the pion mass, the scattering lengths relax to natural values, thus allowing
their determination from the lattice. In fig.~\ref{fig:scattpanel} we exhibit
the $\siii$ scattering length as a function of the pion mass as compared to the
lattice data from Ref.~\cite{fuku}. The left panel has only negligible cutoff
dependence, by construction. The cutoff dependence exhibited in the right panel
demonstrates the incomplete nature of the calculation with implicit pion mass
dependence (the absence of $D_2$). We truncate our curves at $\sim 400~{\rm
  MeV}$ as higher-order effects are increasingly important in this region.

%


\section{Conclusions and Thoughts}

We have addressed several important issues concerning the development of a
consistent and converging EFT to describe multi-nucleon systems.  Chiral
perturbation theory in the single-nucleon sector is a special case of a more
general ordering of operators.  In the presence of more than one nucleon, the
momentum expansion is nonperturbative at some level to accommodate the
fine-tuned scales, while the $m_q$-expansion remains perturbative. In other
words, the perturbative expansion of nuclear forces is an expansion about the
chiral limit. In the $\si$ channel only local operators are treated
nonperturbatively, whereas in the $\siii-\diii$ coupled channels it is
necessary to resum the non-local, singular part of OPE which survives in the
chiral limit.

The amount of chiral-limit physics which must be resummed is determined
phenomenologically channel-by-channel. One can always choose to resum the
entire momentum expansion since perturbative physics remains so even when
treated nonperturbatively.  KSW power counting is the minimal power-counting
scheme to describe systems with unnaturally-large scattering lengths; i.e. one
operator in the momentum expansion is summed to all orders. If the effective
range is also unnaturally large, it too should be resummed and this in turn
constitutes a legitimate ordering of operators.  W power counting is a
prescription which sums a kernel computed to a given order in chiral
perturbation theory to all orders. Here there is a subtlety; one must ensure
that at a given order in the expansion there is a correspondence between
divergences and the local operators which absorb them. This is obeyed by W
power counting in the $\siii-\diii$ coupled channels but not in the $\si$
channel. A bizarre feature that emerges from the nonperturbative nature of the
$\siii-\diii$ coupled channels is the scaling of operators with fractional
powers of the chiral symmetry-breaking scale, e.g. $D_2\sim\lamchi^{-3/2}$

Our partial higher-order calculations in the $\siii-\diii$ coupled channels
suggest that an expansion about the chiral limit will converge. However, a full
NLO calculation is required in order to make a more definite conclusion and to
give meaningful predictions for the deuteron binding energy in the chiral
limit. The use of the square well as a short-distance regulator has proved
valuable in giving analytic formulas for the RG running of the counterterms.
However, the necessity of computing processes with external gauge fields
suggests use of a regulator that manifestly respects gauge invariance, like
dimensional regularization or Pauli-Villars.  An intriguing puzzle remains
concerning the relationship between square-well regularization and the matching
of delta-function interactions to singular potentials. In taking the limit
$R\rightarrow 0$ one finds that the singular potential wavefunctions vanish.
This makes it difficult to understand how a delta-function interaction can
modify the physical asymptotic phase. This puzzle is compounded in the
coupled-channel system.

An EFT with dynamical pions is required to describe both elastic and inelastic
processes involving momenta greater than $\sim\mpi$. For very-low momentum
processes $\ll\mpi$ it is appropriate and consistent to work with $\nopi$.  All
quark-mass dependence is implicit in the coefficients of local operators as
chiral symmetry is not a symmetry of this EFT.

One may worry that observables traditionally used to probe nuclear forces, like
the electric-quadrupole moment, will not be well reproduced by our proposed EFT
and will be cutoff dependent. However contributions from gauge-invariant,
four-nucleon, one-photon operators that arise at higher order in the EFT will
precisely compensate the cutoff dependence and when combined with the one-body
contributions from the nuclear wave function will reproduce these
observables~\cite{KSWem}.

Finally, there is a significant overlap between the approach advocated in this
paper, and a more mechanical approach to the problem.  Haxton and
collaborators~\cite{wick} have shown that the nuclear Shell Model can be
formulated as an effective theory, and its convergence can be dramatically
improved by resumming the ``short-distance'' part of the NN interaction and
then performing perturbation theory with the remaining ``long-distance'' part.
The separation into long- and short-distance is defined numerically by the
shape of the NN potential channel-by-channel, rather than by underlying
hadronic structure. Nonetheless this approach recovers the essential parts of
the expansion we are advocating in this work.

\vspace{1cm}
\noindent

{\large\bf Acknowledgements}

\noindent
We thank W.~Haxton, U.~Mei\ss ner, D.~Phillips, G.~Rupak, S.~Sharpe, D.~Sprung,
R.~Wiringa and M.~Wise for useful discussions.  This research was supported in
part by the DOE grant DE-FG03-97ER41014 (SRB,MJS), and by the Director, Office
of Energy Research, Office of High Energy and Nuclear Physics, Division of
Nuclear Physics, and by the Office of Basic Energy Science, Division of Nuclear
Science, of the U.S.  Department of Energy under Contract No. DE-AC03-76SF00098
(PFB), and by RIKEN, Brookhaven National Laboratory and the U.S. Department of
Energy DE-AC02-98CH10886 (UvK).

\vspace{1cm}
\appendix
\section{The Standard Toy Model of Nuclear Physics: Three Yukawas}

One can gain significant insight into the perturbative structure of the 
$\si$ channel and $\siii-\diii$ coupled channels 
by studying the behavior of a toy theory of 
nucleons interacting in a potential formed from three Yukawa 
exchanges~\cite{KSa}.
The potential has the form
\begin{eqnarray}
V(r) & = & V_s (r) \ +\ V_\pi (r)
\nonumber\\
V_s (r) & = & \alpha_\rho\ { e^{-m_\rho r}\over r}
\  -\ \alpha_\sigma\ { e^{-m_\sigma r}\over r}
\ \ ,\ \ 
V_\pi (r) \ =\ 
-{\bar{\alpha}_\pi}\ { e^{-m_\pi r}\over r}
\ \ \ ,
\label{eq:threeyukpot}
\end{eqnarray}
where the masses
\begin{eqnarray}
m_\rho\ =\ 770\ {\rm MeV}
\ \ ,\ \ 
m_\sigma\ =\ 500\ {\rm MeV}
\ \ ,\ \ 
m_\pi\ =\ 140\ {\rm MeV}
\ \ ,
\label{threeyukmass}
\end{eqnarray}
and couplings
\begin{eqnarray}
\alpha_\rho\ =\ 14.65
\ \ ,\ \ 
\alpha_\sigma\ =\ 7.00
\ \ ,\ \ 
{\bar{\alpha}_\pi}\ =\ 0.075
\ \ ,
\label{threeyukcouplings}
\end{eqnarray}
are chosen, along with a nucleon mass of $M=940~{\rm MeV}$ to
approximately reproduce both the scattering length, and effective
range in the $\si$ channel, $a^{3Y} = -23.7~{\rm fm}$ and $r^{3Y} =
2.57~{\rm fm}$.  In making contact with OPE in the real
world, the coupling ${\bar{\alpha}_\pi}$ depends upon both the pion mass,
$m_\pi$, and the axial coupling constant, $g_A$,
${\bar{\alpha}_\pi}\propto g_A^2 m_\pi^2$.  Therefore, perturbing in
the pion potential is equivalent to perturbing in either $g_A$,
$m_\pi$ or both.  However, it is convenient
to think about expanding about the chiral limit, i.e.  expanding in
$\delta V = V(r;m_\pi) - V(r;0) = V_\pi$.  One significant difference
between this toy theory and the real world is that $M$
does not receive contributions from ``potential pion'' loops, and so
$M=940~{\rm MeV}$ at each order in perturbation theory.  This will
not be the case in the real world, where $M$  depends
analytically and non-analytically on the quark masses.

Solving the Schr\"odinger equation in perturbation theory is straightforward.
The s-wave radial wavefunction, $U_{3Y}^{\rm full} (r)$, satisfies
\begin{eqnarray}
{1\over M} {d^2\over dr^2} U_{3Y}^{\rm full}
\ -\ 
(V-E) U_{3Y}^{\rm full}
& = & 0
\ \ \ .
\label{eq:threeyukfull}
\end{eqnarray}
Writing $U_{3Y}^{\rm full} = U_{3Y}^0 + U_{3Y}^1 + ...$, 
where the superscript denotes the number of insertions of $V_\pi$,
it follows that
\begin{eqnarray}
{1\over M} {d^2\over dr^2} U_{3Y}^0
\ -\ 
(V_s-E) U_{3Y}^0
& = & 0
\nonumber\\
{1\over M} {d^2\over dr^2} U_{3Y}^j
\ -\ 
(V_s-E) U_{3Y}^j
& = & V_\pi U_{3Y}^{j-1}
\ \ \ ,\ \ \ j\ge 1
\ \ \ ,
\label{eq:threeyukpert}
\end{eqnarray}
with boundary conditions at the origin chosen appropriately for each
of the $U_{3Y}^j$.  The s-wave phase-shifts $\delta_{3Y}$ can be found
from the wavefunction at any given order in perturbation theory by
simply projecting out the incoming and outgoing asymptotic plane
waves, e.g. see Ref.~\cite{KSWa}. This method automatically provides a
unitary phase-shift as one is expanding $\exp \left(i 2
\delta\right)$ which necessarily involves $V_\pi$ to all orders.  We
denote this expansion as $E1$.  One may perform a further expansion of
$\delta_{3Y}$ to any given order in $V_\pi$ that is unitary to a given
order in $g_A^2$, and we denote this expansion as $E2$.

\begin{figure}[!ht]
\centerline{{\epsfxsize=3.3in \epsfbox{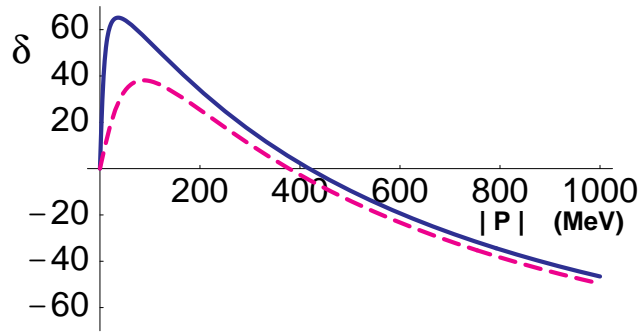}} }
\vskip 0.15in
\noindent
\caption{\it 
The full phase-shift (solid)
and the phase-shift resulting from zero-insertions of
$V_\pi$ (dashed)
in the three-Yukawa toy model.
}
\label{fig:yukLO}
\vskip .2in
\end{figure}

\begin{figure}[!ht]
\centerline{{\epsfxsize=2.8in \epsfbox{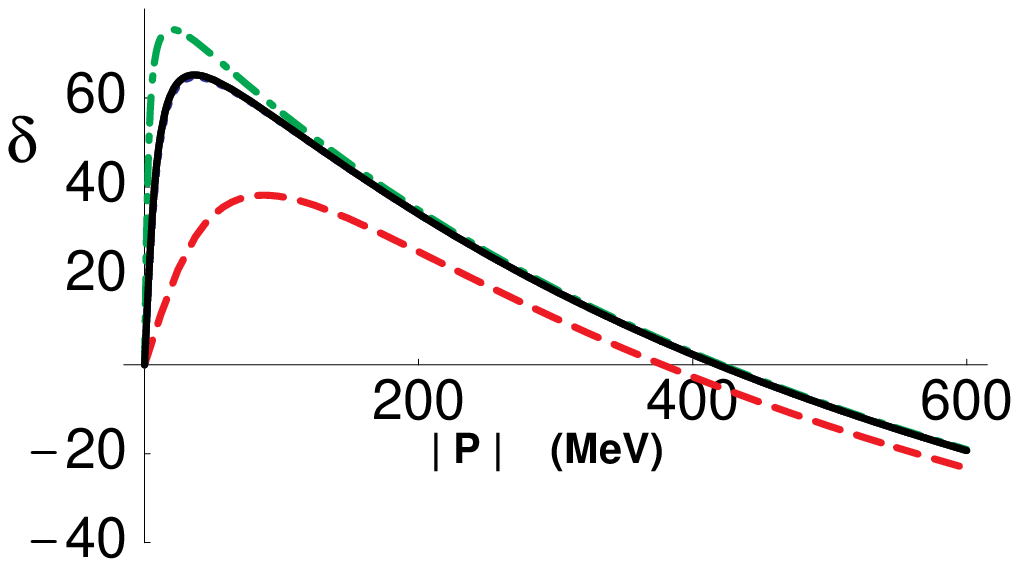}}
{\epsfxsize=2.8in \epsfbox{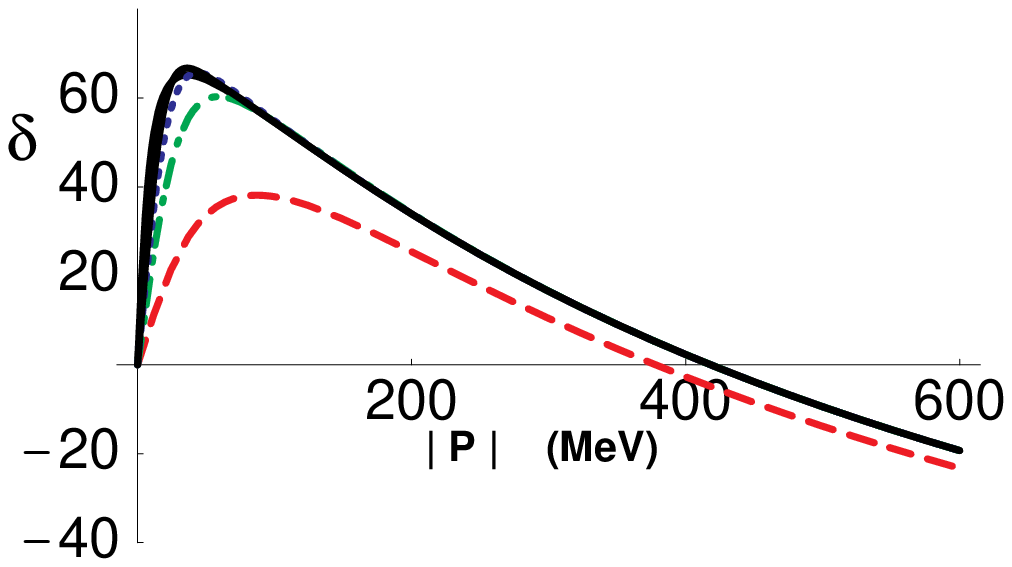}} }
\vskip 0.15in
\noindent
\caption{\it 
The left panel shows the perturbative phase-shifts 
in the three-Yukawa toy model obtained by solving the Schr\"odinger
equation, and  using expansion $E1$, 
while the right panel results from expansion $E2$.
The solid curve is the full phase-shift, while the 
dashed, dot-dashed and dotted curves correspond to 
LO, NLO, and N$^2$LO, respectively.
In both cases the N$^2$LO result essentially 
lies on top of the 
full result.
}
\label{fig:yukrest}
\vskip .2in
\end{figure}

At LO, corresponding to no insertions of $V_\pi$, the 
``short-distance'' phase-shifts
resulting from $E1$ and $E2$ are identical, as shown in fig.~\ref{fig:yukLO}.
%
One sees sizable deviations between the two phase-shifts below
$|{\bf p}|\simle 200~{\rm MeV}$.
The phase-shifts at higher orders in perturbation theory are shown in 
fig.~\ref{fig:yukrest}.
%
It is clear that both expansions 
converge rapidly from $|{\bf p}|\simge 100~{\rm MeV}$.
Further, expansion $E1$ converges rapidly for all $|{\bf p}|$, while
$E2$ converges somewhat more slowly in the very low $|{\bf p}|$ region.

\begin{figure}[!ht]
\centerline{{\epsfxsize=3.3in \epsfbox{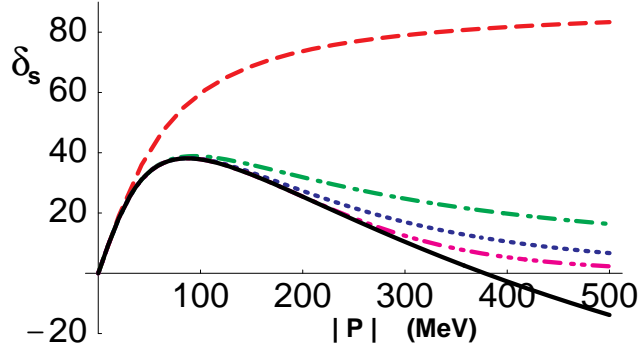}} }
\vskip 0.15in
\noindent
\caption{\it 
Successive effective-range approximations to the 
short-distance phase-shift.
The solid curve is the exact short-distance phase-shift 
while the 
dashed, dot-dashed, dotted and dot-dot-dashed 
curves correspond to the effective-range 
expansion including terms up to $1/a_s$, 
$r_0^s$, $r_1^s$ and $r_2^s$, respectively.
}
\label{fig:yukER}
\vskip .2in
\end{figure}

While we take some joy in demonstrating the perturbative nature of
this theory, it is still the case that we do not know the underlying
short-distance physics responsible for the real-world NN interaction
and somehow it must be extracted from data.  Given that we wish to
keep the perturbative aspect of the problem, we need to most
efficiently describe the short-distance part of the interaction, whose
exact nature we care little about in the low-momentum region.  The
form of this parameterization has been the source of much debate,
i.e. W, KSW or some hybrid.  However, for this particular
problem we know exactly what we can do to parameterize the
short-distance part {\it only}, and that is to use effective range
theory (ER), which should be valid up to momenta of order $|{\bf
p}|\sim m_\sigma/2\sim 250~{\rm MeV}$.  Therefore, we fit a polynomial
of the form
\begin{eqnarray}
|{\bf p}|\cot\delta_s & = & -{1\over a^s}
\ +\ {1\over 2} r^s_0 |{\bf p}|^2
\ +\ r^s_1 |{\bf p}|^4
\ +\ ...
\ \ \ ,
\label{eq:threeyukshort}
\end{eqnarray}
to the ``short-distance'' phase-shift shown in  fig.~\ref{fig:yukLO}, to find
\begin{eqnarray}
a^s & = & -3.38~{\rm fm}
\ \ ,\ \ 
r_0^s\ =\ 2.60~{\rm fm}
\ \ ,\ \ 
r_1^s\ =\ 0.313~{\rm fm}
\ \ ,\ \ 
r_2^s\ =\ 0.156~{\rm fm}
\ \ \ .
\label{eq:threeyukshortfit}
\end{eqnarray}
A comparison between the ``short-distance'' 
phase-shift, and successively better
ER approximations are shown in 
fig.~\ref{fig:yukER}.
%
Retaining only the leading term, i.e. 
$|{\bf p}|\cot\delta_s  =  -{1\over a^s}$ is clearly not such a good
approximation for momenta $|{\bf p}|\simge 40~{\rm MeV}$, and corresponds to
the underlying perturbative scheme of KSW.
However, retaining the short-distance effective range contribution
is  a significant improvement, 
but still appears to begin deviating from the true phase-shift for 
$|{\bf p}|\simge 150~{\rm MeV}$.  However, this is in no way related to 
$m_\pi$, and is a numerical accident.
It is this approximation that corresponds to the counting suggested 
in~\cite{KSa}, and fleshed-out in Ref.~\cite{BScount}.
As one proceeds to higher orders in the ER expansion, the phase
shift is well recovered up to $|{\bf p}| \simge 250~{\rm MeV}$, as 
expected.

%
\begin{figure}[!ht]
\centerline{{\epsfxsize=2.8in \epsfbox{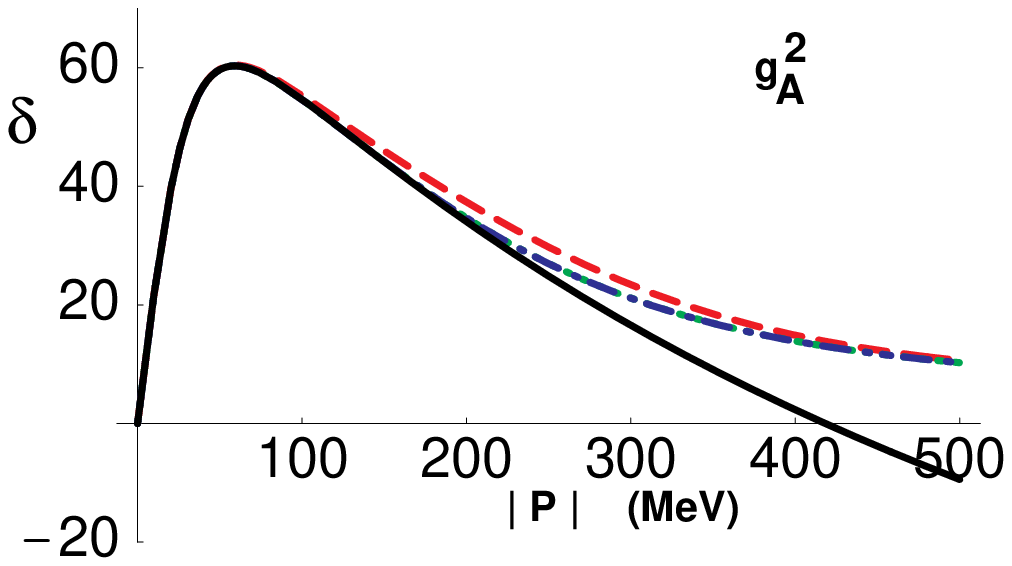}}
{\epsfxsize=2.8in \epsfbox{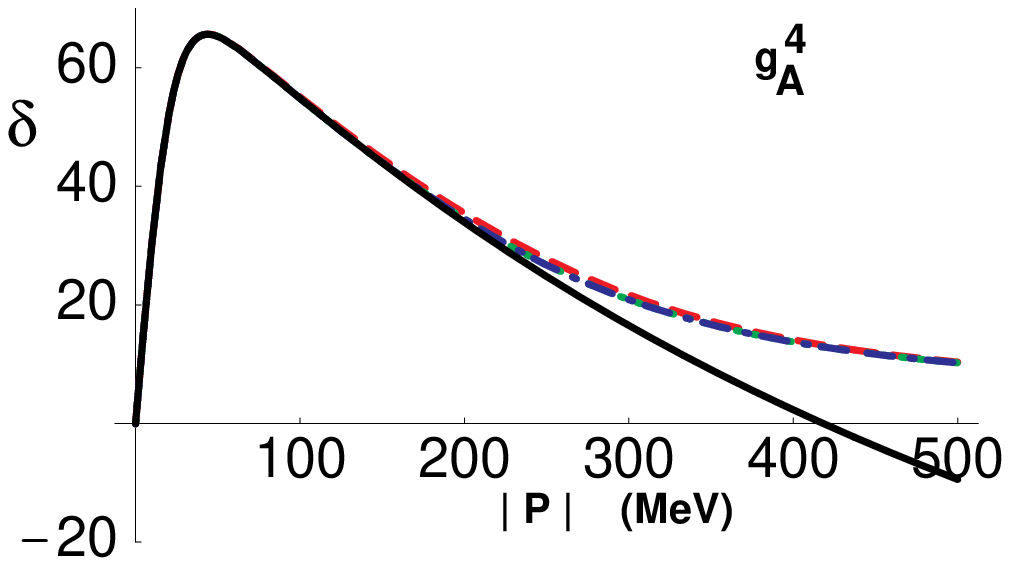}} }
\vskip 0.15in
\noindent
\caption{\it 
Phase-shifts in the EFT of the three-Yukawa toy model.
The dibaryon field describing the short-distance physics
is expanded out to order $|{\bf p}|^6$.
The left panel shows the phase-shift up to one insertion of $V_\pi$.
The solid curve is the full phase-shift at this order
obtained from the Schr\"odinger equation, while
the dashed, dot-dashed and dotted curves correspond to
one perturbative insertion of $V_\pi$
in the EFT expansion including 
the order $g_A^2$ counterterms up to 
$|{\bf p}|^2$, $|{\bf p}|^4$ and $|{\bf p}|^6$,
respectively.
The right panel shows the phase-shift up to two insertions of $V_\pi$.
The solid curve is the full phase-shift at this order
obtained from the Schr\"odinger equation, while
the dashed, dot-dashed and dotted curves correspond to
two perturbative insertions of $V_\pi$ in 
the EFT expansion including the order $g_A^4$ counterterms up to 
$|{\bf p}|^2$, $|{\bf p}|^4$ and $|{\bf p}|^6$,
respectively.
The $g_A^2$ counterterms determined in the left panel 
are used in the ${\cal O}\left(g_A^4\right)$ calculation shown in the 
right panel.
}
\label{fig:yukEFT}
\vskip .2in
\end{figure}
{}From the EFT point of view, we can use a generalized dibaryon field
to include an arbitrary number of terms in the short-distance physics
via $|{\bf p}|\cot\delta_s$.  Further, the long-distance amplitudes
have been computed out to N$^2$LO by Rupak and Shoresh~\cite{RuSh99}.
However, there are additional parameters that need to be determined by
numerical fitting.  They occur because the EFT must be matched to the
full theory.  The short-distance behavior of graphs involving all
three particles, the $\pi,\sigma$ and $\rho$, does not allow for a
clean separation into a dibaryon field defined by the short-distance
parameters of eq.~(\ref{eq:threeyukshortfit}) and pion exchange.
However, one can perform the separation into a dibaryon field defined
by the short-distance parameters of eq.~(\ref{eq:threeyukshortfit}),
pion exchange and local finite counterterms of the form
$g_A^{2k}\left(m_\pi/m_\sigma\right)^n , g_A^{2k}\left(|{\bf
p}|/m_\sigma\right)^m,...$.  The LO amplitude in the EFT consists
solely of a dibaryon field with the parameters of
eq.~(\ref{eq:threeyukshortfit}).  The NLO amplitude in the EFT
consists of the dibaryon field, a single insertion of $V_\pi$ and both
momentum-independent and momentum-dependent counterterms.  As
expected, the fit improves as the number of counterterms is increased.
It is important to note that the size of the counterterms is
determined by $m_\sigma$. However, if the $V_\pi$ insertion is {\it
not} included and the counterterms are refit, one finds that their
size is now set by $m_\pi$, and the overall quality of the fit is
noticeably degraded.  At N$^2$LO the pattern continues with the same
dibaryon field, two insertions of $V_\pi$, and more counterterms.  The
phase-shifts at various orders in $V_\pi$ are shown in
fig.~\ref{fig:yukEFT}.

\end{document}